\documentclass[a4paper,fleqn,usenatbib]{mnras}

\usepackage[T1]{fontenc}
\usepackage{ae,aecompl}

\usepackage{graphicx}	
\usepackage{amsmath}	
\usepackage{amssymb}	

\title[Complex evolutionary paths of local infrared bright
galaxies]{The complex evolutionary paths of local infrared bright
  galaxies: a high angular resolution mid-infrared view}

\author[A. Alonso-Herrero et al.]
{A. Alonso-Herrero,$^{1,2,3}$\thanks{E-mail: aalonso@cab.inta-csic.es} 
R. Poulton,$^{4}$
P. F. Roche,$^{2}$ 
A. Hern\'an-Caballero,$^{5}$
\newauthor
I. Aretxaga,$^{6}$
M. Mart\'{\i}nez-Paredes,$^{6,7}$ C. Ramos Almeida,$^{8,9}$ 
M. Pereira-Santaella,$^{10}$
\newauthor
T. D\'{\i}az-Santos,$^{11}$ 
N. A. Levenson,$^{12}$ C. Packham,$^{3,13}$ 
 L. Colina,$^{10}$ P. Esquej,$^{14,15}$
\newauthor
 O. Gonz\'alez-Mart\'{\i}n,$^{7}$ 
 K. Ichikawa,$^{13}$  M. Imanishi,$^{16,13,17}$  
 J. M. Rodr\'{\i}guez Espinosa,$^{8,9}$ 
\newauthor
 C. Telesco$^{18}$\\
$^{1}$Centro de
  Astrobiolog\'{\i}a (CAB, CSIC-INTA), ESAC Campus, E-28692 Villanueva de la
Ca\~nada, Madrid, Spain\\
$^{2}$Department of Physics, University of Oxford, Oxford OX1 3RH,
UK\\
$^{3}$Department of Physics and Astronomy, University of Texas at San
Antonio, San Antonio, TX 78249, USA\\
$^{4}$Department of Physics and Astronomy, University of Sussex, Brighton
BN1 9QH, UK\\
$^{5}$Departamento de Astrof\'{\i}sica, Facultad de CC. F\'{\i}sicas, Universidad Complutense de Madrid, 28040 Madrid, Spain\\
$^{6}$Instituto Nacional de Astrof\'{\i}sica, Optica y Electr\'onica
  (INAOE), 72000 Puebla, Mexico\\
$^{7}$Centro de Radioastronom\'{\i}a y Astrof\'{\i}sica (CRyA-UNAM), 3-72 (Xangari),
8701, Morelia, Mexico\\
$^8$Instituto de Astrof\'{\i}sica de Canarias (IAC), E-38205 La Laguna,
Tenerife, Spain\\
$^9$Departamento de Astrof\'{\i}sica, Universidad de la Laguna (ULL),
E-38206 La Laguna, Tenerife, Spain\\ 
$^{10}$Centro de
  Astrobiolog\'{\i}a (CAB, CSIC-INTA), 28850 Torrej\'on de Ardoz,
  Madrid, Spain\\
$^{11}$N\'ucleo de Astronom\'{\i}a de la Facultad de Ingenier\'{\i}a, Universidad Diego Portales, Av. Ej\'ercito Libertador 441, Santiago, Chile\\
$^{12}$Gemini Observatory, Casilla 603, La Serena, Chile\\
$^{13}$National Astronomical Observatory of Japan, 2-21-1 Osawa, Mitaka,
Tokyo 181-8588, Japan\\
$^{14}$European Space Astronomy Centre (ESAC)/ESA, E-28691, Villanueva de la Ca\~nada, Madrid, Spain\\
$^{15}$ISDEFE, Beatriz de Bobadilla 3, 28040 Madrid, Spain\\
$^{16}$Subaru Telescope, 650 North A'ohoku Place, Hilo, Hawaii, 96720,
USA\\
$^{17}$Department of Astronomical Science,
The Graduate University for Advanced Studies (SOKENDAI),
Mitaka, Tokyo 181-8588, Japan\\
$^{18}$Department of Astronomy, University of Florida, Gainesville, FL 32611, USA
}
\vspace{-0.6cm}
\date{Accepted XXX. Received YYY; in original form ZZZ}

\pubyear{2016}

\begin{document}
\label{firstpage}
\pagerange{\pageref{firstpage}--\pageref{lastpage}}
\maketitle

\begin{abstract}
  We investigate the evolutionary connection between
local IR-bright galaxies ($\log L_{\rm IR}\ge 11.4\,L_\odot$)
  and   quasars. We use  
 high angular resolution ($\sim$ 0.3-0.4\,arcsec$\sim $ few hundred parsecs) 
$8-13\,\mu$m ground-based spectroscopy to disentangle the AGN
mid-IR properties from those of star formation. 
The comparison between the nuclear $11.3\,\mu$m PAH feature 
emission and that measured with {\it Spitzer}/IRS indicates that the 
star formation is extended over a few kpc in the IR-bright galaxies. The AGN contribution to the 
total IR luminosity of IR-bright galaxies is  lower than in
quasars. 
Although  the dust distribution is predicted to change as  IR-bright galaxies
evolve to IR-bright quasars and then to optical
quasars, we show that the AGN mid-IR emission of all the quasars in
our sample  is not significantly different. In contrast, the nuclear
emission of IR-bright galaxies with low AGN contributions appears more heavily embedded in dust although there is no
clear trend with the interaction stage or projected nuclear separation. This suggests
that the changes in the distribution of the nuclear obscuring 
material may be taking place rapidly and at
different interaction stages washing out the evidence of an evolutionary path. 
When compared to normal AGN, the nuclear star formation activity of quasars
appears to be dimming whereas it is enhanced in some IR-bright nuclei, suggesting that the latter are in an
earlier star-formation dominated phase.
\end{abstract}




\section{Introduction}\label{sec:intro}

The evolutionary connection between ``cool'' ({\it IRAS} $25\,\mu$m
to $60\,\mu$m colors $f_{25}/f_{60}\lesssim 0.3$) ultraluminous infrared (IR)
galaxies (ULIRGs, defined to have $8-1000\,\mu$m IR luminosities $L_{\rm
  IR}>10^{12}\,L_\odot$), to ``warm'' ($f_{25}/f_{60}\gtrsim 0.3$) ULIRGs, and
optical 
quasars was proposed almost thirty years ago \citep{Sanders1988}. In this
scenario IR bright activity and in particular
ULIRG activity is triggered by interactions between gas-rich
galaxies that merge, go through an optical quasar phase and eventually
evolve into elliptical galaxies
\citep[see review by][]{SandersMirabel1996}.
Numerical simulations predict that during the merger phase both intense star
formation and a dust enshrouded active galactic nucleus (AGN) phase
co-exist in the dusty nuclear regions of
most  ULIRGs before energetic feedback from an AGN and/or star
formation clears the nuclear region of gas and dust
\citep[see e.g.][]{DiMatteo2005,Hopkins2008}. Plenty of observations now support
this scenario at least in terms of the 
AGN detection rate and increased AGN bolometric contribution to the system luminosity  with increasing IR luminosity for local luminous infrared galaxies (LIRGs,
with IR luminosities $L_{\rm IR}=10^{11}-10^{12}\,L_\odot$) and ULIRGs
\citep{Kim1995, Veilleux1995, Veilleux2009ULIRG, Imanishi2007, Yuan2010,  Nardini2008, Nardini2010, Petric2011, AAH2012}.

There has been considerable effort devoted to understanding the
relation between the IR luminosity and the 
morphology, merger state, star formation  rate and AGN dominance of
ULIRGs and
how all these properties relate to those of optical quasars. In
particular, using {\it Spitzer} Infrared Spectrograph \citep[IRS,
][]{Houck2004} mid-infrared (mid-IR) 
spectroscopic observations,\cite{Veilleux2009ULIRG} showed that ULIRGs
do not seem to follow a simple evolutionary path to optical quasars, either in terms
of the AGN bolometric dominance or the Eddington ratios. Moreover,
\cite{Farrah2009} based on the same data set proposed that the lifecycle
of local ULIRGs consists of three phases with the first one during
pre-coalescence being dominated by star formation activity. The
subsequent final evolution of the ULIRG may or may not go through a
quasar phase depending on the properties of the progenitor galaxies.  
For quasars, \cite{Haas2003} proposed an evolutionary
sequence which is driven by the geometry of the
dust in their nuclear regions to explain the variety of observed optical and IR spectral indices.

\begin{table*}
	\centering
	\caption{The sample of local IR-bright galaxies.}
	\label{tab:sampleULIRG}
	\begin{tabular}{lccccccccc} 
		\hline

Galaxy         &IRAS Name      &Redshift &Dist &Class      &$\log L_{\rm IR}$   &Morph &Ref  &Other name\\
               &               &         &(Mpc)  &        &($L_\odot$)   &     &  &\\
\hline
IZw1           &IRAS00509+1225 &0.058900 &248  &Sy1        &11.86 &IVb   &E06  &PG0050+124\\
Mrk~1014        &IRAS01572+0009 &0.163110 &748  &Sy1        &12.53 &IVb   &Y10  &PG0157+001\\ 
NGC~1614        &IRAS04315$-$0840 &0.015938 &65.5 &Cp         &11.58 &IIIb    &Y10\\
Mrk~1073        &IRAS03117+4151 &0.023343 &95.3 &Sy2        &11.39 & Pair/IIIa?  &V95\\
IRAS~08572+3915 &-              &0.058350 &254  &NW/SE-Sy2: &12.11 &IIIb  &Y10\\
UGC~5101        &IRAS09320+6134 &0.039367 &168  &Sy2:       &12.00 &IV    &Y10\\
Arp~299         &IRAS11257+5850 &0.010411 &45.2 &Sy2/L      &11.83 &IIIb  &GM06\\
Mrk~231         &IRAS12540+5708 &0.042170 &181  &Sy1        &12.50 &IVb   &Y10  &UGC08058\\
IRAS~13349+2438 &      -        &0.107641 &483  &Sy1        &12.32 &IV/V?     &W09  &[HB89]1334+246 \\
Mrk~463         &IRAS13536+1836 &0.050355 &219  &E-Sy1/W-Sy2        &11.78 &IIIb  &GM07\\
IRAS~14348$-$1447  &       -      &0.083000 &366  &SW/NE-Cp:  &12.26 &IIIb  &Y10\\ 
Mrk~478         &IRAS14400+3539 &0.079055 &347  &Sy1        &11.52 &V     &E06  &PG1440+356\\ 
NGC~6240        &IRAS16504+0228 &0.024480 &103  &L          &11.83 &IIIb  &Y10\\
IRAS~17208$-$0014 &        -      &0.042810 &181  &HII       &12.40 &IV    &Y10\\
\hline
	\end{tabular}

        Notes. --- Dist is the luminosity distance taken from NED for $H_0 = 73\,{\rm km \,s}^{-1}\,{\rm Mpc}^{-1}$, $\Omega_M = 0.27$, and $\Omega_\lambda = 0.73$.  The IR luminosity is in the $8-1000\,\mu$m range. The morphology
class follows the criteria defined by \cite{Veilleux2002}:
IIIa wide binary (nuclear separation $ >$ 10\,kpc); IIIb close binary (nuclear separation $<$ 10\,kpc);
IVa: diffuse merger; IVb: compact merger; V: old merger;  Pair: In a pair.
The references are for the optical class
except for Mrk~478 and IZw1 which is for the IR luminosity: GM06: \cite{GarciaMarin2006}, Y10: \cite{Yuan2010},
W09: \cite{Wu2009},  GM07: \cite{GarciaMarin2007}, V95: \cite{Veilleux1995}, E06: \cite{Evans2006}.

\end{table*}

\begin{table*}
\begin{minipage}{10cm}
\begin{center}
	\caption{The comparison sample of IR-weak quasars.}
	\label{tab:sampleQSO}
	\begin{tabular}{lccccccccc} 
		\hline

Galaxy         &Redshift &Dist &Class      &$\log L_{\rm IR}$   &
Other name\\
                          &         &(Mpc)  &        &($L_\odot$)   &  \\
		\hline
Mrk~335     & 0.025785   & 103  &Sy1   &10.75  & PG0003+199\\  
PG0804+761 & 0.100000 & 443 &Sy1   &11.83  &\\    
PG0844+349 & 0.064000 & 279 &Sy1   &10.94  &\\    
PG1211+143 & 0.080900 & 358 &Sy1   &11.61  & \\    
3C273          &0.158339 &734  &Sy1        &12.68 & PG1226+023\\
PG1229+204 & 0.063010 & 276 &Sy1   &10.90  & \\    
PG1411+442 & 0.089600 & 396 &Sy1   &11.61  & \\    
Mrk~1383    & 0.086570 & 383 &Sy1   &11.55  & PG1426+015 \\   
Mrk~841     & 0.036422 & 157 &Sy1.5 &10.96  & PG1501+106\\    
Mrk~509     & 0.034397 & 141 &Sy1.5 &11.17  & \\    
		\hline
	\end{tabular}

Notes.---The IR luminosities are from \cite{Weedman2012} corrected to our cosmology
except for Mrk~841 which is computed from the {\it IRAS} fluxes listed in NED. 
\end{center}
\end{minipage}
\end{table*}

In spite of the high AGN incidence in the population of local
IR-bright galaxies,  little is known about the detailed 
properties of the dust surrounding the AGN 
and star formation activity of their  nuclear regions and
how they relate to the putative torus of the Unified Model. 
Studies of local universe type 1 quasars
have shown that their unresolved near- and mid-IR emission can be reproduced with
clumpy dusty torus models with small covering factors \citep{Mor2009,
  MartinezParedes2016,Mateos2016}, whereas quasars optically
classified as type 2 have larger covering factors \citep{Mateos2016}.
On the other hand, the nuclear mid-IR AGN 
emission of local ULIRGs has only been modeled for a handful of objects. The results vary from source to source with some 
being modeled with
high covering factor torus models and high foreground extinctions
\citep{AAH2013, Mori2014, MartinezParedes2015},  others requiring fully
embedded heating sources \citep{Levenson2007}, and some having relatively
small covering factors and properties similar to those of other
Seyfert 1 nuclei \citep[for instance the nucleus of
the LIRG NGC~7469, see][]{Hoenig2010, AAH2011, Ichikawa2015}.

In this work we take advantage of the nearly factor of 10 improved
angular resolution
afforded by mid-IR instruments on 8-10\,m class telescopes when compared
to {\it Spitzer}/IRS to investigate the evolutionary connection
between local IR-bright galaxies and quasars.  We use new and existing high angular resolution ($\sim 0.3-0.4\,$arcsec)
ground-based mid-IR spectroscopy mostly obtained with the 
CanariCam \citep{Telesco2003, Packham2005}
instrument on the 10.4\,m Gran Telescopio CANARIAS (GTC).
By using high angular resolution we are able to isolate 
nuclear scales of a few hundred parsecs where the AGN and star
formation processes are believed to be more
tightly coupled \citep{HopkinsQuataert2010}. 
We observed a sample of local LIRGs and ULIRGs with 
IR luminosities $\log L_{\rm IR}\ge 11.4\,L_\odot$, as at these
luminosities most systems are interacting galaxies or mergers
\citep[see e.g.][and references therein]{Hung2014, Larson2016}. We
dub these systems IR-bright galaxies. As a comparison
sample we observed optical quasars which are mostly Palomar-Green (PG)
quasars from the Bright Quasar Sample \citep{Schmidt1983}. Among the
IR-bright galaxy sample there are four IR-bright PG quasars (see Section~\ref{sec:sample}
for more details on our definition). These quasars 
tend to show more pronounced merger-induced morphological anomalies
and thus might be
at an earlier evolutionary stage than IR-faint
quasars \citep{Veilleux2009QSO}. \cite{Cao2008} reached a similar
conclusion by
comparing the properties of the mid-IR {\it Spitzer}/IRS spectra of
IR-bright and IR-weak quasars. Therefore, IR-bright quasars
might represent a transition phase to the evolution to {\it classical}
optical quasars.

The paper is organized as follows. Section~\ref{sec:sample} describes the
samples of  IR-bright galaxies and  quasars.
Section~\ref{sec:observations} gives details on the new and existing
mid-IR observations used in this
work. Section~\ref{sec:analysis} presents the analysis of the mid-IR spectroscopy, whereas
Section~\ref{sec:nuclearmidIRemission}
discusses the mid-IR AGN and nuclear star formation properties of local IR-bright
galaxies and
the comparison sample of quasars. In Section~\ref{sec:evolution} we investigate the possible
evolutionary connection between the mid-IR AGN properties and nuclear
star formation activity  of IR-bright
galaxies and quasars and finally in Section~\ref{sec:conclusions} we
present our conclusions.
Throughout this work we use the following cosmology:  $H_0 = 73\,{\rm km \,s}^{-1}\,{\rm Mpc}^{-1}$,
$\Omega_M = 0.27$, and $\Omega_\lambda = 0.73$.

\section{The samples}\label{sec:sample}
Our sample of IR-bright galaxies contains 14 local systems with IR
luminosities in the range
$\log L_{\rm IR} = 11.4-12.5\,L_\odot$  (see
Fig.~\ref{fig:histogramLIR} and Table~\ref{tab:sampleULIRG}), 
hosting an AGN (see below) and 
with available high angular resolution ($\sim$0.3-0.4\,arcsec)
mid-IR imaging and spectroscopy (see
Section~\ref{sec:observations}).  All but one are part of our 
GTC/CanariCam mid-IR survey of local AGN \citep[see][for details]{AAH2016}.
According to their IR luminosities, 7 systems are LIRGs and 7 are ULIRGs.
We selected our sample of IR-bright
galaxies to cover mostly close (projected nuclear distances of $<10\,$kpc) interaction and merger phases.  Using the morphological classification
of \cite{Veilleux2002} we have (see
Table~\ref{tab:sampleULIRG}) 6 close interacting galaxies termed
class IIIb, 5 mergers termed class IV,  and 1 old merger termed class
V. Mrk~1073 is part of the Perseus cluster and probably paired with 
UGC~2612 \citep{Levenson2001} at a projected distance greater than
10\,kpc, so likely in the IIIa class.
Finally, the
ULIRG/IR-bright quasar  IRAS~13349+2438 \citep[see][and below]{Low1989} has no
morphological classification but it shows a compact
appearance which probably indicates a merger The projected nuclear separations
between the nuclei of galaxies classified as IIIb are as follows, 6.1\,kpc for IRAS~08572+3915,
5.5\,kpc  for IRAS~14348$-$1447, 4.5\,kpc  for Arp~299, 3.8\,kpc for Mrk~463
\citep{GarciaMarin2009}, 3\,kpc for NGC~1614 \citep[this galaxy is
classified as a minor merger by][and references therein]{Larson2016},
0.7\,kpc for NGC~6240 \citep{Komossa2003}, 
and 0.2\,kpc for IRAS~17208$-$0014 \citep{Medling2014, Medling2015}.

The high angular resolution
($0.3-0.4\,$arcsec) of our mid-IR data (see
Sections~\ref{sec:newdata} and \ref{sec:olddata}) allows us to resolve the emission from the
individual nuclei of those IR-bright galaxies in our sample in class IIIb except
for IRAS~17208$-$0014   (see also Section~\ref{sec:Spitzer}). Of the
systems with double nuclei, only two
systems have IR-bright nuclei for which we could obtain sufficiently
high signal-to-noise (S/N) ratio spectra (see Section~\ref{sec:observations}),
namely for the eastern and western components of
Arp~299 \citep[IC~694 and NGC~3690, respectively, see][and Fig.~\ref{fig:Arp299}]{AAH2013} and
NGC~6240N and NGC~6240S \citep[see][and
Fig.~\ref{fig:NGC1614,NGC6240,IRAS17208}]{AAH2014}. For the other
three close interacting 
systems in our sample only one nucleus is sufficiently IR-bright to
obtain ground-based mid-IR spectroscopy (see next section), namely, IRAS~08572+3915N,
Mrk~463E, and IRAS~14348$-$1447S \citep[see][for more
details]{AAH2016}. For the minor merger NGC~1614 we only observed the
main galaxy.
Therefore, our sample of local IR-bright galaxies contains 16
individual nuclei.

In terms of the optical spectral classification, all the IR-bright
galaxy sample nuclei but one are
classified as AGN, that is, Seyfert (Sy), LINER (L), or Composite (Cp, that
is, located between the H\,{\sc ii} and Seyfert classes in optical
line ratio diagnostic diagrams). The only exception is  
IRAS~17208$-$0014, which is classified as H\,{\sc ii} in the optical but has 
evidence of AGN activity based on X-ray, IR and
submillimeter data
\citep{GonzalezMartin2009,Veilleux2009ULIRG,GarciaBurillo2015}. 
The median distance of the IR-bright galaxy sample is 181\,Mpc. 

Our ground-based mid-IR spectroscopic observations were by
necessity flux-limited \citep[mid-IR fluxes above approximately 20\,mJy within small
    apertures, see][]{AAH2016} and required compact morphologies.
This may introduce some biases in the
results derived for our sample of IR-bright galaxies. In particular,
  our sample of IR-bright galaxies (also the sample of quasars, see
  below)  might be biased towards more AGN-dominated sources and/or sources with compact
  nuclear star formation. For instance, those nuclei in the interacting systems
  for which we could not get observations are classified as Sy or Cp
  but were both
  faint and not compact. This means that  the
AGN likely contribute little to their nuclear mid-IR emission. Therefore 
these AGN have very low fractional contributions to the total IR
luminosities of the systems (see Section~\ref{sec:AGNlum}).
Finally, we note that \cite{DiazSantos2011} showed that local LIRGs and ULIRGs
  with a significant AGN contribution tend to show relatively compact
  mid-IR continuum
  emission. Thus it is likely that our sample of IR-bright galaxies
  with AGN classifications is not too different from the general population of local
  IR-bright galaxies with significant AGN contributions.

We also compiled mid-IR spectroscopic
observations with a similar high angular resolution for a comparison sample of 10 optically selected
local quasars, mostly   PG
quasars taken from the work of \cite{MartinezParedes2016}. These
quasars were chosen to match bolometrically the luminosities 
of our sample of IR-bright galaxies \citep[see Figure~1 of][for the
$2-10\,$keV luminosities of the PG quasars]{AAH2016}.
We list the properties of the sample of quasars in Table~\ref{tab:sampleQSO}.
The median luminosity distance of the  quasar sample is
319\,Mpc.  

Although the IR luminosities of six of the 10 optically selected
quasars would put them in the  (U)LIRG class (see
Table~\ref{tab:sampleQSO} and Fig.~\ref{fig:histogramLIR}),
they are IR-weak quasars \citep[see][]{Low1989}, 
based on their IR to optical $B$-band luminosity
ratios (typically ratios $L_{\rm IR}/L_B$ between 1  and 3, as
calculated using total $B$-band magnitudes from NED).  
Among the  IR-bright galaxies in our sample there are also
four optically selected quasars (IZw1, Mrk~1014,
Mrk~478,  and IRAS~13349+2439), of which three are in the PG
catalog (see 
Table~\ref{tab:sampleULIRG}). These are, however IR-bright quasars based on
their observed IR to $B$-band ratios ($L_{\rm
  IR}/L_B$ ratios  between 4 and 11). Our IR-bright and IR-weak quasar
classifications are consistent with
the far-IR strong and weak classes of \cite{Netzer2007} based on the observed
$60\,\mu$m to $15\,\mu$m ratios of PG quasars.
Finally, Mrk~231 is also considered the
nearest quasar \citep{Boksenberg1977} and has a dusty broad absorption
line region  obscuring the continuum source and the standard broad absorption
line region \citep{Veilleux2013}. Thus, Mrk~231 might
be a bona-fide
emerging IR-bright quasar in the local universe.

\begin{figure}
\hspace{-0.5cm}
\resizebox{1.1\hsize}{!}{\rotatebox[]{0}{\includegraphics{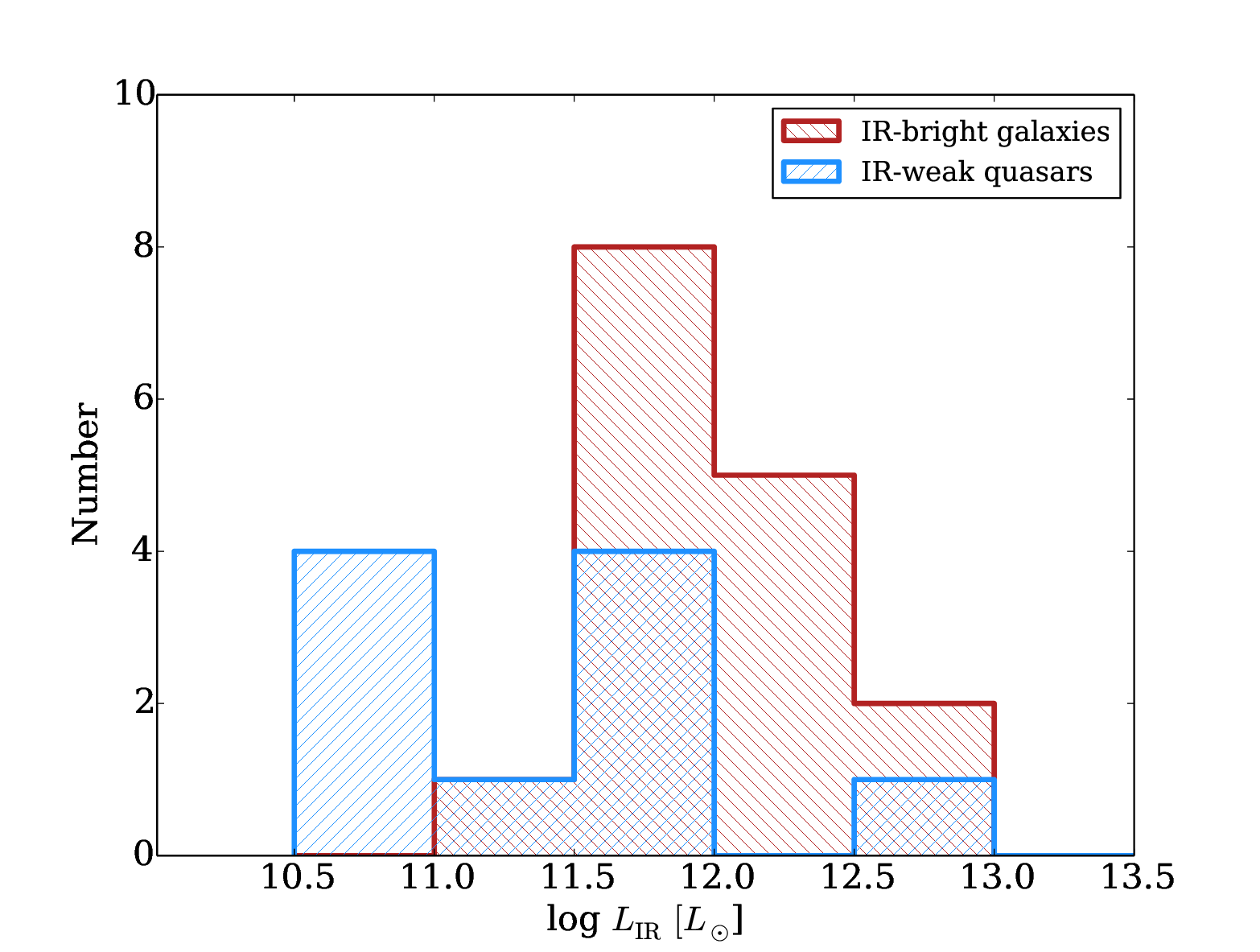}}}
    \caption{Distributions of the IR luminosities of the systems for
      the sample of IR-bright galaxies (red histogram) and the comparison sample of
     IR-weak quasars (blue histogram).} 
    \label{fig:histogramLIR}
\end{figure}

\begin{figure*}
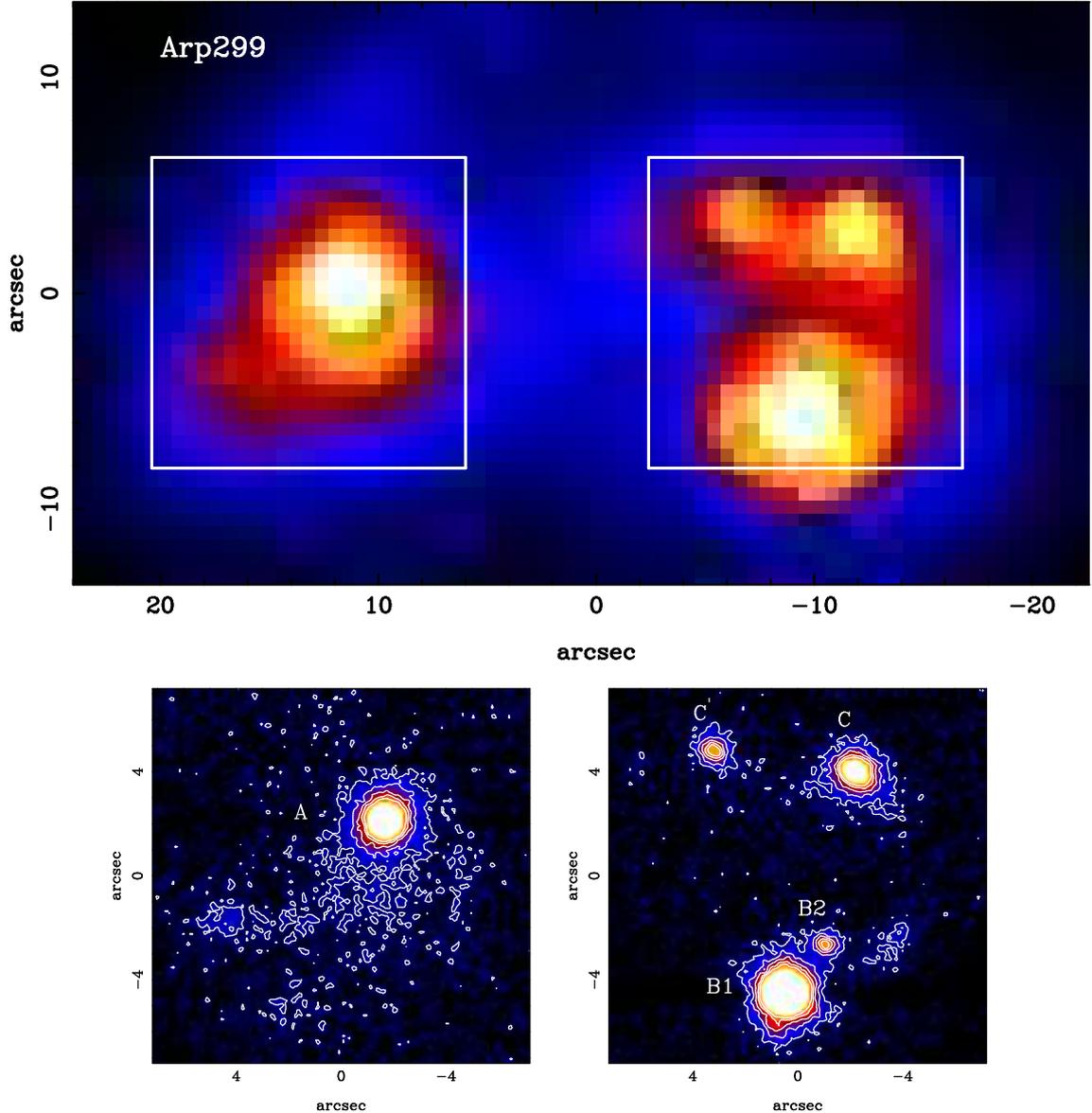


\resizebox{0.85\hsize}{!}{\rotatebox[]{0}{\includegraphics{figure2.ps}}}

\vspace{0.3cm}
\resizebox{0.33\hsize}{!}{\rotatebox[]{-90}{\includegraphics{figure2a.ps}}}
\hspace{0.3cm}
\resizebox{0.33\hsize}{!}{\rotatebox[]{-90}{\includegraphics{figure2b.ps}}}
\caption{The upper panel is the
{\it Spitzer}/IRAC $8\,\mu$m image of Arp~299 shown in a square root scale to
emphasize the diffuse emission. The white squares
represent the approximate FoV of the  GTC/CanariCam images shown in the
lower panels of the two galaxies in the system: IC~694
  or Arp~299A (left), and NGC~3690 (right). The nuclear region of NGC~3690 is the
  B1 source. B2 is a bright optical source, whereas C and C' are
  bright star forming regions in the overlapping region of the two
  galaxies. The CanariCam images were obtained with  the Si-2 filter
  ($\lambda_{\rm c}=8.7\,\mu$m). We smoothed the CanariCam images using a
Gaussian function with $\sigma =0.6$ pixels. The orientation of the
images is north up, east to
the left, and are shown in a linear scale. } 
    \label{fig:Arp299}
\end{figure*}

\begin{figure*}
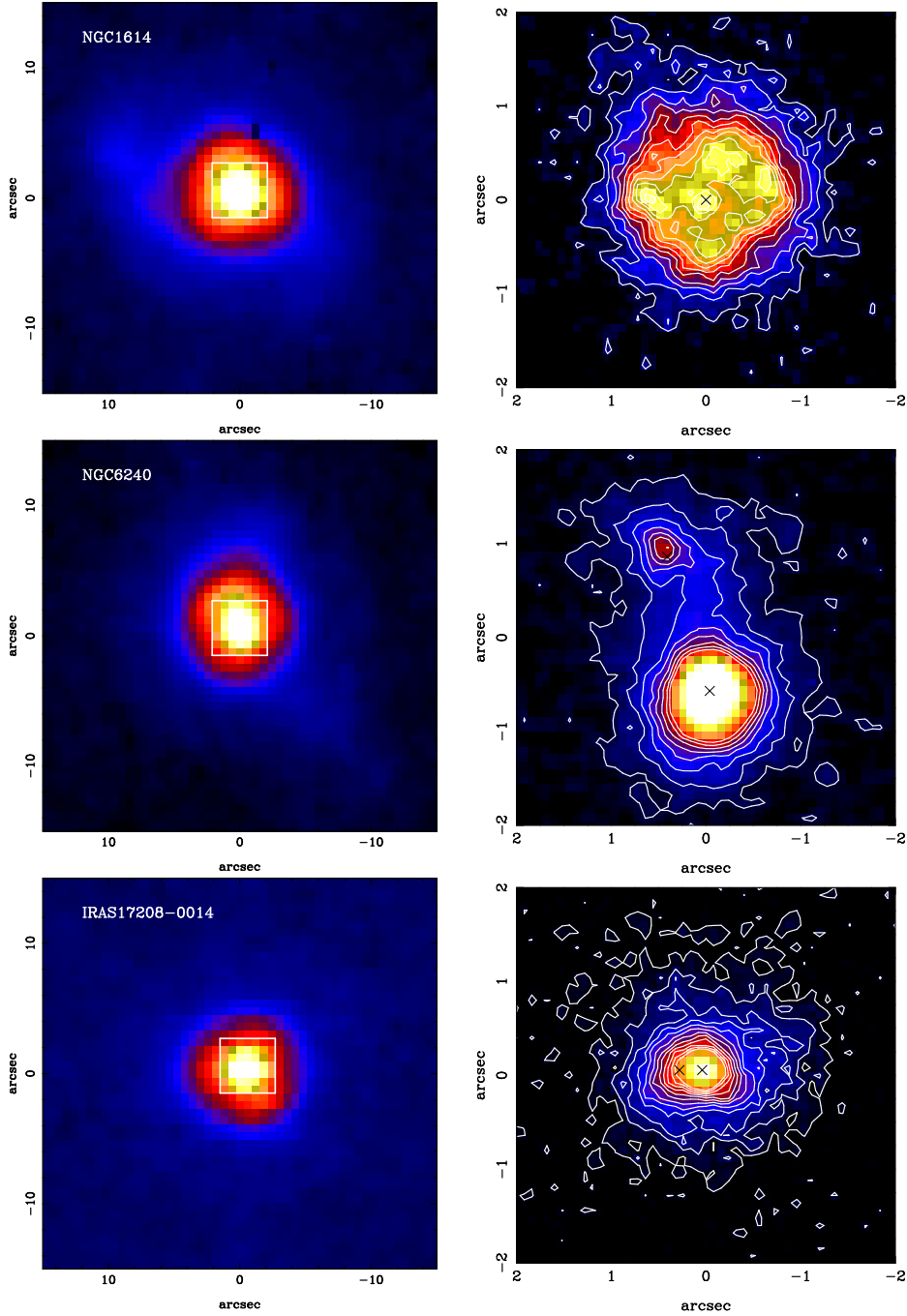

\resizebox{0.33\hsize}{!}{\rotatebox[]{0}{\includegraphics{figure3e.ps}}}
\hspace{0.3cm}
\resizebox{0.33\hsize}{!}{\rotatebox[]{-90}{\includegraphics{figure3d.ps}}}
\resizebox{0.33\hsize}{!}{\rotatebox[]{0}{\includegraphics{figure3c.ps}}}
\hspace{0.3cm}
\resizebox{0.33\hsize}{!}{\rotatebox[]{-90}{\includegraphics{figure3f.ps}}}
\resizebox{0.33\hsize}{!}{\rotatebox[]{0}{\includegraphics{figure3g.ps}}}
\hspace{0.3cm}
\resizebox{0.33\hsize}{!}{\rotatebox[]{-90}{\includegraphics{figure3h.ps}}}
\caption{{\it Top panels.} GTC/CanariCam 
  image (right, shown in a linear scale) at $8.7\,\mu$m of the central
  region of NGC~1614,
  which 
clearly resolves the nucleus (marked with a black cross) and the circumnuclear
  ring of star formation \citep[see][for a
  detailed study of this galaxy]{PereiraSantaella2015}. The
  {\it Spitzer}/IRAC $8\,\mu$m image (left, shown in a square root scale) shows a larger FoV with the white square
  representing the approximate FoV of the right panel and also the approximate size
of the {\it Spitzer}/IRS SL slit width (3.7\,arcsec).  The orientation
of the images is north up, east to the left.
{\it Middle panels.} Same as upper panels but for NGC~6240. The GTC/CanariCam $8.7\,\mu$m image (right) of the
  nuclear region of  NGC~6240 clearly resolves the two nuclei of this
  interacting system \citep[see][for more
    details]{AAH2014, AAH2016}.
{\it Bottom panels.} Same as the upper panel but for IRAS~17208$-$0014. The
nearly nuclear point source seen in the IRAC $8\,\mu$m image (left) is clearly resolved in the
GTC/CanariCam image (right). We tentatively identify the western nucleus with the
peak of the $8.7\,\mu$m emission. The  projected separation of the two nuclei \citep[$\sim
0.24\,$arcsec $\sim$ CanariCam 3 pixels,
see][]{Medling2014,Medling2015}, shown as black crosses, is slightly smaller than the FWHM of the
CanariCam image \citep[$0.26\,$arcsec, see ][]{AAH2014}.} 
    \label{fig:NGC1614,NGC6240,IRAS17208}
\end{figure*}

\section{Observations}\label{sec:observations}

\subsection{New GTC/CanariCam imaging observations}\label{sec:newdata}
We obtained new mid-IR high angular resolution imaging observations of the
interacting LIRG Arp~299 with GTC/CanariCam on the night of
2016 January 26 in queue mode. These observations were part of our GTC/CanariCam AGN guaranteed
time program (PI: C. Telesco). We used  the Si-2 filter which has a central
wavelength of $\lambda_{\rm c}= 8.7\,\mu$m   and a width of $\Delta
\lambda_{\rm cut} = 1.1\,\mu$m, at 50\% cut-on/off. 
The CanariCam $320\,{\rm pixel} \times 240\,{\rm pixel}$ Si:As detector has a plate scale 
of $0.0798\,$arcsec pixel$^{-1}$ which provides a field of view (FoV) of 
$26\,{\rm arcsec} \times 19\,{\rm arcsec}$. We therefore required two
pointings to cover the nuclei and bright star forming regions of Arp~299.
The mid-IR spectroscopy of
the bright nuclei of Arp~299 was presented in \cite{AAH2013}. 

The observations were taken using the standard mid-IR chop-nod
technique with on-source integration times of 608\,s for each of the
two pointings. We also
obtained an image of a standard star with the same filter to perform the photometric
calibration and assess the image quality of the observations. 
We reduced the data using the CanariCam pipeline {\sc redcan}
\citep{GonzalezMartin2013} that includes for imaging observations stacking of the individual
observations, rejection of bad images and flux calibration.  Figure~\ref{fig:Arp299}
shows the fully reduced image of the two galaxies of Arp~299. Using
the image of the standard star we measured an image quality of the
observations of 0.48\,arcsec (FWHM). The GTC/CanariCam Si-2 filter imaging data of the rest of the
IR-bright galaxies in the sample are presented in \cite{AAH2016}
except for IZw1 which is in the sample of optically selected quasars
analyzed by \cite{MartinezParedes2016}.

\subsection{Existing high angular resolution mid-IR spectroscopic observations}\label{sec:olddata}

We compiled high angular resolution (image quality of $\sim 0.3-0.4\,$arcsec, FWHM)
mid-IR $\sim 7.5-13\,\mu$m spectroscopy of
the sample of  14 local IR-bright systems listed in
Table~\ref{tab:sampleULIRG} and the comparison sample of 10 IR-weak
quasars in Table~\ref{tab:sampleQSO}.  These data were obtained with
ground-based instruments on $8-10\,$m class telescopes.
The majority of systems in both samples were observed with
the CanariCam instrument on the GTC
in spectroscopic mode with a spectral resolution of $R=\lambda/\Delta
\lambda \sim 175$ as part of the 
CanariCam AGN guaranteed time program, an ESO/GTC large program
\citep[ID: 182.B-2005][]{AAH2016} 
or through the GTC Mexican open time \cite[see][]{MartinezParedes2016}.
The ULIRG IZw1 and the PG quasar 
Mrk~509 were observed in the mid-IR using VISIR \citep{Lagage2004} on the VLT by 
\cite{Burtscher2013} and \cite{Hoenig2010}, respectively, with a spectral resolution
of $R \sim 300$  and similar image quality to the CanariCam spectra. 
We reduced all the CanariCam imaging and spectroscopic data using the {\sc redcan} pipeline \citep{GonzalezMartin2013}.
We refer the reader to \cite{AAH2016} and \cite{MartinezParedes2016}
for full details on the CanariCam observations, including the
extraction of the spectra, the correction for possible slit losses,
and their photometric
calibration.  

The GTC/CanariCam and VLT/VISIR mid-IR spectroscopic observations were taken with similarly sized slit widths 
  of 0.52\,arcsec and 0.75\,arcsec, respectively. At the median distances of the IR-bright
galaxy sample  and the IR-weak quasar sample (see Section~\ref{sec:sample}), the physical sizes
probed by our ground-based
slits range between 112\,pc and 1396\,pc, and between 248\,pc and
1376\,pc, respectively.

\subsection{Archival {\it Spitzer}/IRAC and IRS observations}\label{sec:Spitzer}

For both samples we obtained fully reduced {\it Spitzer}/IRS low spectral
resolution ($R\sim 60-120$) spectra from the Cornell
Atlas of Spitzer/IRS Sources \citep[CASSIS, ][version LR7]{Lebouteiller2011},
except for the nuclear regions of Arp~299 for which we used the
spectra of \cite{AAH2009}. In this work we only used the spectral
range covered by the short-low (SL) mode, $\sim 5-15\,\mu$m, as this
is the range used by the spectral decomposition tool (see
Section~\ref{sec:deblendIRS})  and the 
spectra with optimal extraction regions. The slit width of the SL
spectra is 3.7\,arcsec.

We also retrieved from the {\it Spitzer} archive
fully calibrated Infrared Array Camera \citep[IRAC,][]{Fazio2004}
images at $8\,\mu$m rebinned to a pixel size of 0.6\,arcsec. The FWHM
of the IRAC
images is approximately 2\,arcsec. In Figs.~\ref{fig:Arp299} and \ref{fig:NGC1614,NGC6240,IRAS17208} we
show a few examples in the IR-bright galaxy sample comparing the {\it Spitzer}/IRAC
$8\,\mu$m morphologies and the nuclear morphologies resolved by our
GTC/CanariCam $8.7\,\mu$m images. These figures emphasize the gain in
angular resolution of the ground-based mid-IR observations. 

\begin{table*}
\scriptsize
\caption{{\sc deblendIRS} results for   the
  ground-based high angular resolution
  spectra of the IR-bright galaxy sample.}
\label{tab:deblendIRS_ULIRG}
\begin{tabular}{@{}lcccccccccc@{}}
\hline
Galaxy & $\chi^2$ & AGN $\nu L_{\nu} (12\,\mu{\rm m})$  &
\multicolumn{3}{c}{Mid-IR Contribution} & AGN Frac. at $12\mu$m   &  
AGN $S_{\rm Sil}$ & AGN $\alpha_{\rm MIR}$ & $L_{\rm IR}({\rm
                                             AGN})/L_{\rm  IR}$\\
 & & (${\rm erg\,s}^{-1}$) & AGN & PAH & STR\\
\hline
IZw1       & 2.2 &  $6.5 \times 10^{44}$ &
0.82 & 0.00 & 0.18& 0.87[0.78, 0.96] & 0.20[ 0.05, 0.39] &-1.68[-2.13,-1.29] & 0.37\\ 
Mrk~1014   & 4.9 &  $9.4 \times 10^{44} (^*)$ &
0.90 & 0.10 & 0.00& 0.90[0.85, 0.97] & 0.22[0.00, 0.52]  &-1.22[-1.86,-0.81] & 0.12\\ 
Mrk~1073   & 2.0 &  $2.2 \times 10^{43}$ &
0.61 & 0.15 & 0.24& 0.71[0.54, 0.87] &-0.96[-1.59,-0.33] &-2.00[-2.73,-1.28] & 0.05\\ 
NGC~1614    & 7.6   &$2.0 \times 10^{43}$  &
0.61 & 0.39 & 0.00& 0.63[0.57, 0.72] &-0.75[-1.07,-0.53] &-2.94[-3.75,-2.55] & 0.03\\ 
IRAS~08572+3914N &65.2 &$3.5 \times 10^{44}$  &
1.00 & 0.00 & 0.00& 0.99[0.98, 1.00] &-4.16[-4.31,-4.04] &-0.14[-0.27,-0.02] & 0.16\\ 
UGC~5101   & 5.3 &$4.6 \times 10^{43}$  &
0.82 & 0.18 & 0.00& 0.76[0.55, 0.89] &-2.42[-3.50,-1.39] &-1.41[-2.14,-0.78] & 0.03\\ 
IC~694      &12.3 & $2.2 \times 10^{43}$ &
0.78 & 0.20 & 0.02& 0.86[0.82, 0.91] &-3.46[-3.62,-3.34] &-1.66[-1.79,-1.54] & 0.02\\ 
NGC~3690B1  & 2.1 & $7.3 \times 10^{43}$ &
0.65 & 0.00 & 0.35& 0.79[0.76, 0.82] &-1.67[-1.89,-1.41] &-1.89[-2.05,-1.71] & 0.06\\ 
Mrk~231    & 4.6 & $1.7 \times 10^{45}$  &
0.90 & 0.06 & 0.04& 0.93[0.89, 0.97] &-0.68[-0.81,-0.54] &-2.20[-2.56,-1.96] & 0.22\\ 
IRAS~13349+2438 & 2.8 &$3.2 \times 10^{45}(^*)$  &
0.99 & 0.01 & 0.00& 0.97[0.94, 0.99] & 0.17[-0.01, 0.34] &-0.75[-1.23,-0.37] & 0.65\\ 
Mrk~463E   & 2.5 & $6.8 \times 10^{44}$ &
0.92 & 0.08 & 0.01& 0.87[0.78, 0.96] &-0.47[-0.72,-0.22] &-2.36[-2.77,-1.96] & 0.68\\ 
IRAS~14348$-$1447S & 3.4 & $4.6 \times 10^{43}(^*)$  &
0.57 & 0.39 & 0.03& 0.54[0.24, 0.83] &-1.55[-3.62,-0.09] &-1.34[-2.55,-0.20] & 0.02\\ 
Mrk~478    & 1.8 & $2.9 \times 10^{44} (^*)$ &
0.89 & 0.11 & 0.00& 0.76[0.58, 0.91] & 0.04[-0.30, 0.37] &-1.70[-2.47,-1.07] & 0.37\\ 
NGC~6240N  & 4.9 & $1.0 \times 10^{43}$ &
0.42 & 0.58 & 0.00& 0.42[0.15, 0.62] &-1.32[-3.51,-0.03] &-1.58[-2.64,-0.39] & 0.01\\ 
NGC~6240S  &15.6 &$7.7 \times 10^{43}$  &
0.58 & 0.39 & 0.04& 0.66[0.60, 0.72] &-1.82[-3.49,-1.03] &-2.16[-3.77,-1.61] & 0.07\\ 
IRAS17208$-$0014& 3.0 & $1.5 \times 10^{43}$ &
0.31 & 0.69 & 0.00& 0.42[0.17, 0.64] &-1.68[-3.69,-0.13] &-1.34[-2.51,-0.20] & 0.01\\ 

\hline
\end{tabular}

Notes.--- The listed $\chi^2$ values are reduced ones. For the AGN
rest-frame $12\,\mu$m luminosities, those galaxies marked with $(^*)$ the
ground-based spectra do not cover the rest-frame $12\,\mu$m and the
values were estimated from the extrapolation of the AGN best-fit
template.  The mid-IR contributions of the AGN, PAH, and STR
components are 
in the $5-15\,\mu$m spectral range and within the slit.
For the fitted AGN fractional contribution at $12\,\mu$m
within the slit, mid-IR $5-15\,\mu$m spectral index and strength of
the silicate feature we give the median value and in 
brackets the 16\% and 84\% percentiles of the PDFs.
\end{table*}

\begin{table*}
\scriptsize
\caption{{\sc deblendIRS} results for the
  ground-based high angular resolution
  spectra of the IR-weak quasars.}
\label{tab:deblendIRS_QSO}
\begin{tabular}{@{}lcccccccccc@{}}
\hline
Galaxy & $\chi^2$  & AGN $\nu L_{\nu} (12\,\mu{\rm m})$  &
\multicolumn{3}{c}{Mid-IR Contribution} & AGN Frac. at $12\mu$m   & 
AGN $S_{\rm Sil}$ & AGN $\alpha_{\rm MIR}$  & $L_{\rm IR}({\rm
                                             AGN})/L_{\rm  IR}$\\
 & & (${\rm erg\,s}^{-1}$)& AGN & PAH & STR\\
\hline
Mrk~335    & 2.0& $4.3 \times 10^{43}$  &
 0.78& 0.02& 0.19& 0.58[0.34, 0.82] &-0.09[-0.97, 0.38] &-1.58[-2.56,-0.73] & 0.32\\  
 PG0804+761& 2.7& $9.2 \times 10^{44}(^*)$ &
 0.82& 0.00& 0.18& 0.90[0.85, 0.99] & 0.24[ 0.09, 0.40] &-1.40[-1.81,-0.96] & 0.56\\  
 PG0844+349& 3.6& $1.3 \times 10^{44}$  &
 1.00& 0.00& 0.00& 0.88[0.79, 0.96] & 0.27[ 0.00, 0.50] &-1.76[-2.55,-1.28] & 0.60\\  
 PG1211+143& 2.8& $5.8 \times 10^{44}(^*)$ &
 0.95& 0.02& 0.03& 0.87[0.76, 0.96] & 0.25[ 0.04, 0.46] &-1.51[-1.91,-1.13] & 0.58\\  
 3C273     & 1.2& $4.7 \times 10^{45}(^*)$  &
 0.91& 0.05& 0.04& 0.92[0.86, 0.97] & 0.09[-0.09, 0.27] &-0.93[-1.25,-0.58] & 0.41\\  
PG1229+204& 3.2& $1.3 \times 10^{44}$  &
 0.93& 0.07& 0.00& 0.82[0.69, 0.94] & 0.11[-0.26, 0.43] &-1.78[-2.58,-1.13] & 0.65\\  
 PG1411+442& 1.4& $4.4 \times 10^{44}(^*)$  &
 0.85& 0.00& 0.15& 0.79[0.64, 0.93] & 0.11[-0.13, 0.36] &-1.34[-1.84,-0.80] & 0.44\\  
  Mrk~1383   & 3.1&$5.3 \times 10^{44}(^*)$  &
 0.99& 0.00& 0.01& 0.87[0.75, 0.96] & 0.14[-0.13, 0.38] &-1.87[-2.66,-1.41] & 0.62\\ 
 Mrk~841    & 2.6 & $1.5 \times 10^{44}$&
 0.94& 0.00& 0.06& 0.92[0.83, 0.97] &-0.13[-0.34, 0.14] &-2.26[-2.70,-1.86] & 0.70\\  
 Mrk~509    & 0.9& $1.9 \times 10^{44}$ &
 0.89& 0.02& 0.09& 0.86[0.71, 0.95] & 0.06[-0.21, 0.29] &-1.83[-2.58,-1.38] & 0.52\\  
 
\hline
\end{tabular}

Notes.---  All the columns are as in Table~\ref{tab:deblendIRS_ULIRG}.

\end{table*}

The GTC/CanariCam $8.7\,\mu$m images of Arp~299 show emission stemming not only from the
bright mid-IR sources (A, B1, B2, C and C') but also from H\,{\sc ii}
regions in the spiral
arms of the eastern component and to the north west of B1, as well as
extended emission around B1 \citep[see Fig.~\ref{fig:Arp299} and][] {AAH2009}.  The IRAC $8\,\mu$m
image of  IRAS~17208$-$0014  shows a bright nearly point source, which
is clearly resolved and extended in the east-west direction in the GTC/CanariCam $8.7\,\mu$m (see
Fig.~\ref{fig:NGC1614,NGC6240,IRAS17208}, bottom panels). The projected
separation of the two nuclei identified in this galaxy
\citep[see][]{Medling2014, Medling2015} is
approximately the same as the FWHM of the image (0.26\,arcsec) and
cannot therefore be resolved. However, based on the nuclear Br$\gamma$
\citep{Medling2014} and Pa$\alpha$ \citep{PiquerasLopez2012}
morphologies of this system and the similarity between hydrogen
recombination line and $8.7\,\mu$m morphologies of local LIRGs
\citep{DiazSantos2008}, we tentatively identify the peak of the CanariCam emission
with the western nucleus of
IRAS~17208$-$0014. \cite{GarciaBurillo2015} proposed that this
nucleus hosts an obscured \citep[i.e.,
Compton-thick, see][]{GonzalezMartin2009} AGN  which is responsible for the  energetic 
outflow detected in molecular gas in this system. We note  that
the extraction aperture used for the GTC/CanariCam spectra includes
emission from the two nuclei. NGC~1614 and NGC~6240 were discussed in
detail in \cite{PereiraSantaella2015} and \cite{AAH2014}, respectively.

\section{Analysis}\label{sec:analysis}

\subsection{Deriving the  AGN mid-IR properties}\label{sec:deblendIRS}
Even on sub-arcsecond resolution the mid-IR emission of local (U)LIRG
nuclei can have a significant contribution from star formation
activity, as revealed by
the detection of polycyclic
aromatic hydrocarbon (PAH) features, in particular  the $11.3\,\mu$m
PAH feature \citep[see e.g.][]{Soifer2002,
  DiazSantos2010, Mori2014, PereiraSantaella2015, AAH2014, AAH2016}. To study the AGN mid-IR emission of local
(U)LIRG nuclei, we need to disentangle it from that due to star
formation. Spectral decomposition methods have been proven effective
in doing so for local (U)LIRGs \citep[see e.g.][]{Nardini2008, Nardini2010,
  AAH2012, MartinezParedes2015} and have been mostly applied to {\it
  Spitzer}/IRS spectra.

We use the {\sc
  deblendIRS} tool\footnote{http://denebola.org/ahc/deblendIRS}
\citep{HernanCaballero2015} to do the spectral decomposition of the
ground-based mid-IR spectra of the two samples. {\sc deblendIRS} is an {\sc idl/gdl}-based routine
that uses  {\it Spitzer}/IRS observational spectral templates to decompose the
mid-IR $5-15\,\mu$m spectra of galaxies into AGN, interstellar (PAH), and stellar (STR)
components. Although {\sc deblendIRS} was originally designed to be used
with IRS spectra, it can be easily adapted to do the spectral
decomposition of ground-based spectroscopy by setting the adequate
spectral range. We note, however that this routine uses templates taken with IRS. We
  are therefore assuming that the mid-IR emission associated with star formation activity
  on kpc scales resembles that taking place in the vicinity of the AGN.

{\sc deblendIRS} provides the best-fitting model to the data and it also gives
the probability
distribution functions (PDF) of the fitted AGN, PAH and STR fractional components to the
$5-15\,\mu$m emission within the slit as
well as of AGN mid-IR properties such as the mid-IR spectral index
$\alpha_{\rm MIR}$ in the range $\sim
5-15\,\mu$m, the strength of the $9.7\,\mu$m silicate feature $S_{\rm 
  Sil}$ (defined as $\ln f/f_{\rm c}$ where $f$ is the flux density at
the feature and $f_c$ is the flux density of the underlying
continuum). We calculated the  AGN rest-frame  $12\,\mu$m
monochromatic luminosity, $\nu L_\nu(12\mu{\rm m})$, 
using the best-fit AGN component at that wavelength. 

In 
Appendix~\ref{appendix} we show two examples of the graphical output
from {\sc deblendIRS} for an IR-bright galaxy with AGN-dominated
mid-IR emission (Mrk~463E)
and an IR-bright galaxy  with a star formation-dominated nuclear
mid-IR spectrum (IRAS~17208$-$0014).  
In Tables~\ref{tab:deblendIRS_ULIRG} and ~\ref{tab:deblendIRS_QSO} we
list the results from the {\sc deblendIRS} spectral decomposition
of the nuclear spectra of the IR-bright galaxies and comparison quasars, respectively. We
provide for each galaxy the reduced $\chi^2$ value of the best fitting model,
the rest-frame $12\,\mu$m monochromatic AGN luminosity, 
the best fit value of the AGN, PAH and stellar contributions in the
$5-15\,\mu$m spectral range within the slit, the median value
of the AGN fraction contribution at rest-frame $12\,\mu$m and
1$\sigma$ confidence interval (i.e., the 16\% and 84\% percentiles of
the PDF), the median value of the AGN mid-IR spectral index and
1$\sigma$ confidence interval, and the median value of the strength of
the silicate feature (positive values indicate that the feature is in
emission and negative values in absorption) and 1$\sigma$ confidence
interval. Since the AGN spectral templates are flatter than those
accounting for the stellar and interstellar emission \citep[i.e., STR and PAH templates, see Figure~2
of][]{HernanCaballero2015}, the
contribution from the AGN at long wavelengths is generally lower than the 
integrated values for the entire $5-15\,\mu$m range.

The reduced $\chi^2$ values derived from the fits to the mid-IR
ground-based spectra of the IR-bright galaxies and quasars in our sample 
indicate that in the majority of the
cases the reliability of the fits is good. The highest value of $\chi^2$ is for 
 IRAS~08572+3915N for which none of the AGN templates, including the ``obscured
 AGN'' templates, reproduces its
 extremely deep silicate absorption. We refer the reader
   to Appendix~\ref{appendix} for a more detailed discussion on fits of
   sources with deep silicate features and high values of $\chi^2$.
Deep silicate features are believed to be related to additional
extinction due to either dust lanes in the host
galaxy or the merger process
\citep[see][and also Section~\ref{sec:obscuringmaterial}]{Goulding2012, GonzalezMartin2013}.
We finally note
that for sources with $S_{\rm Sil}<-2$ both the AGN fraction and AGN
luminosity should be considered as upper limits, 
because the "obscured AGN" templates could be composite sources.

Finally, we also performed the spectral decomposition of the {\it Spitzer}/IRS spectra
of the IR-bright galaxy and quasar samples. In Appendix~2 we show the
comparisons between 
the derived AGN spectral index and strength of the silicate feature for the IRS and
ground-based spectra using {\sc deblendIRS}. We demonstrate that the results are compatible for
the majority of the sources. We however  use the {\sc deblendIRS}
AGN results obtained from the ground-based mid-IR spectra because they allow us to compare
them with the star formation
activity on nuclear scales (see Sections~\ref{sec:measurePAH} and \ref{sec:AGNvsSF}).

\subsection{Measuring the $11.3\,\mu$m PAH feature}\label{sec:measurePAH}
The PAH features and in particular the $11.3\,\mu$m PAH emission are
good proxies for deriving the star formation rate (SFR) in starburst
galaxies,  LIRGs and ULIRGs
\citep{Brandl2006,Farrah2007,AAH2013_LIRGS}
as well as in Seyfert galaxies and other AGN
\citep{Shi2007,Netzer2007,DiamondStanic2012}. 
To measure the $11.3\,\mu$m PAH feature from both
the high angular resolution and the {\it Spitzer}/IRS
spectra, we fitted a local continuum
on both sides of the feature (rest frame $10.75-11.0\,\mu$m and $11.65-11.9\,\mu$m) and
integrated the feature within the spectral range 
$\lambda_{\rm rest}=11.05-11.55\,\mu$m.
We refer the reader to \cite{HernanCaballero2011} for more details on this method and
to \cite{Esquej2014} for its implementation for ground-based mid-IR spectroscopy.
We estimated the uncertainties in the flux and equivalent width (EW) of
the feature by performing
Monte Carlo simulations using the calculated dispersion
around the measured fluxes and EWs in 100 simulations of the
original spectrum. For the CanariCam spectra we used the random noise distributed as the 
standard deviation calculated over 5 pixels. For the 
IRS spectra we only used the rms error  provided by CASSIS. 


In Table~\ref{tab:measurePAH} we give the luminosity and EW of the $11.3\,\mu$m
PAH feature measured on nuclear scales (given in parsecs) and the ratios
between the nuclear and the
{\it Spitzer}/IRS fluxes for both the local IR-bright galaxies and the
IR-weak quasars.
We note that for galaxies at approximately $z>0.08$, the spectral range covered by
the ground-based data does not allow us to measure this PAH feature
and therefore are not listed in this table.  As we did in \cite{Esquej2014}, we
  considered a detection when the flux of the $11.3\,\mu$m PAH feature
  was at least twice its error. This is equivalent to having the PAH feature
  detected with a significance of $2\sigma$ or
higher (i.e., with a 95\% confidence or higher). 
For the nuclear non-detections of the PAH feature the upper limits to the $11.3\,\mu$m PAH
luminosity and EW in the table are given at the $2\sigma$ level. For
IZw1, Mrk~463E,  Mrk~335, and PG1229+204 nuclear $11.3\,\mu$m PAH
emission may be present but 
with a low ($\sim 1-1.5\sigma$) significance. Thus for these four nuclei in
Table~\ref{tab:measurePAH} we give the measurement
value plus 1$\sigma$ uncertainty as upper limits. 


\begin{table*}
\caption{High angular resolution nuclear measurements of the
  $11.3\,\mu$m PAH feature and nuclear SFRs.}
\label{tab:measurePAH}
\begin{tabular}{@{}lccccc@{}}
\hline
Galaxy & Region & $L(11.3\mu{\rm m\,PAH})$ & EW$(11.3\mu{\rm m\,PAH})$ & Nuc/IRS & Nuclear SFR \\
       & (pc)  & (10$^{42}\,{\rm erg\,s}^{-1}$) & ($\mu$m) & ratio & ($M_\odot \,
                                                        {\rm yr}^{-1})$\\
\hline
\multicolumn{6}{c}{IR-bright galaxies}\\
\hline
IZw1          & $\le 810$            & $<0.6$ & $< 0.01$       & $<$0.5 & $<$3\\
NGC~1614 & $160\times610$ & 0.76   & $0.39\pm 0.02$ & 0.40 & 0.8 \\
Mrk~1073  & $\le 230$            & 0.17   & $0.08\pm 0.01$ & 0.11 & 0.9\\
IRAS~08572+3915N &$\le 570$ & < 0.4&< 0.02&--  & -- \\
UGC~5101  & $392\times 755$ & 0.60   & $0.21\pm 0.03$ & 0.19 & 3.0\\
IC~694        & $\le 112$            & 0.11   & $0.10\pm 0.01$ & 0.08 & 0.6\\
NGC~3690B1   & $\le 112$      & $<0.1$ & $< 0.01$       & $<$0.1 & $<$0.4\\
Mrk~231     & $\le 420$          & $<1.2$ & $< 0.01$       & $<$1 & $<$6\\
Mrk~463E   & $\le 499$           & $<0.8$ & $< 0.01$       & $<$0.5 & $<$4\\
Mrk~478    & $\le 750$ & 1.53 & $0.05\pm 0.01$ & 0.53 & 7.6\\
NGC~6240N  & $247\times 475$  & 0.70   & $0.87\pm 0.13$ & 0.19 & 3.6\\
NGC~6240S  & $\le 247$          & 1.53   & $0.27\pm 0.02$ & 0.41 & 7.6\\
IRAS~17208$-$0014 & $421 \times 809$ &1.19   & $0.56\pm 0.08$ & 0.26 &
  6.0\\
\hline
\multicolumn{6}{c}{IR-weak quasars}\\
\hline
Mrk~335         & $\le 248$    & $<0.1$ & $<$0.03 & $<$1 & --\\
PG~0844+349 & $\le 621$      & $<0.4$ & $<$0.04 & -- & --\\
PG~1229+204 & $\le 615$      & $<0.5$& $<$0.04 & $<$3 & --\\
Mrk~841    & $\le 367$      & $<0.2$ & $<$0.01 & $<$3 & --\\
Mrk~509    & $\le 480$      & 0.19   & $0.011\pm 0.004$ & 0.20 & 1.0\\
\hline
\end{tabular}

Notes.--- The {\it Spitzer}/IRS SL slit covers both nuclei of
NGC~6240, Mrk~463, and IRAS~08572+3915,
whereas for Arp~299 the two nuclei, NGC~3690B1 and IC~694, were
observed separately (see Alonso-Herrero et al. 2009). The upper limits
to the size of the emitting region indicate that the mid-IR nuclear emission
appeared unresolved at the observed angular resolution of the
ground-based spectroscopy (see
Alonso-Herrero et al. 2016) and correspond to the slit width. The
nuclear to IRS ratios refer to the luminosity ratios. The
nuclear SFR take into account the factor of 2 needed to correct the
PAH luminosities measured with a local continuum to the values
obtained with {\sc pahfit} and use the \cite{DiamondStanic2012} calibration.
\end{table*}

\section{Nuclear mid-IR emission of quasars and IR-bright galaxies}\label{sec:nuclearmidIRemission}
\subsection{AGN rest-frame $12\,\mu$m luminosities and contribution to IR luminosity}\label{sec:AGNlum}
\begin{figure}
	\includegraphics[width=1.05\columnwidth]{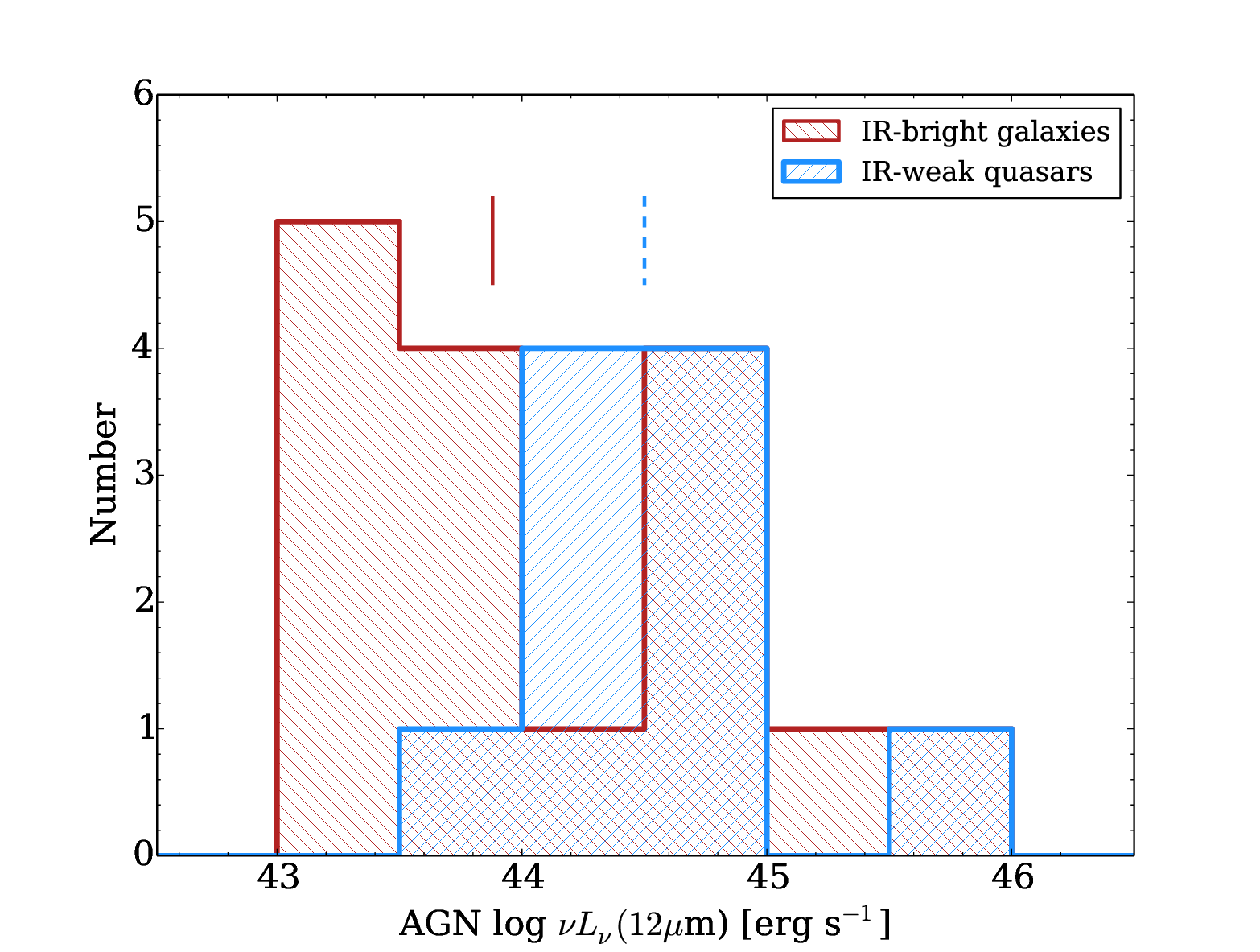}
    \caption{Distributions of the AGN rest-frame $12\,\mu$m monochromatic 
      luminosities derived with {\sc deblendIRS} for the the sample of
      IR-bright galaxies (red histogram) and the comparison sample of 
      IR-weak quasars (blue solid histogram). The vertical
      lines mark the median of the distributions for the IR-bright
      galaxies (solid red line) and IR-weak quasars (blue dashed
    line).} 
    \label{fig:histogramL12muAGN}
\end{figure}

Using the results from {\sc deblendIRS} we derived the monochromatic
rest-frame $12\,\mu$m luminosities of the AGN in the samples of
the IR-bright galaxies and IR-weak quasars. For those nuclei where the observed 
ground-based $\sim 7.5-13\,\mu$m spectra did not cover the rest-frame
$12\,\mu$m (marked in Tables~\ref{tab:deblendIRS_ULIRG} and \ref{tab:deblendIRS_QSO}), we derived the luminosities by extrapolating the
best-fit AGN component. Tables~\ref{tab:deblendIRS_ULIRG} and \ref{tab:deblendIRS_QSO}
list the derived rest-frame $12\,\mu$m AGN luminosities for the
IR-bright galaxy and quasar samples,
respectively, and Fig.~\ref{fig:histogramL12muAGN} shows the corresponding histograms.

As can be seen from this figure, the median values of the AGN luminosities \citep[using the
  $12\,\mu$m AGN luminosity as a proxy, see e.g.][and references therein]{Asmus2014, Asmus2015} of the local
IR-bright galaxies are slightly lower than those of the IR-weak quasars. However, the
IR-bright galaxy AGN show
a broader distribution. Their faint luminosity end  ($\log \nu L_\nu(12\mu{\rm m}) \sim 10^{43}\,{\rm erg \,s}^{-1}$)
still corresponds to typical values of local Seyfert galaxies
\citep{Asmus2014, AAH2016} even for those nuclei optically classified
as L, Cp and H\,{\sc ii} (see Table~\ref{tab:sampleULIRG}).
The high end of the $12\,\mu$m AGN luminosity distribution
($\log \nu L_\nu(12\mu{\rm m}) \ge 10^{45}\,{\rm erg \,s}^{-1}$) of
the sample of local IR-bright galaxies is
contributed by  two
IR-bright quasars (e.g., Mrk~1014 and IRAS~13349+2438) as well as
the emerging quasar Mrk~231.  Based on the observed distribution of
the $12\,\mu$m AGN luminosities of the PG quasars in our sample, we consider that
AGN with monochromatic $12\,\mu$m luminosities greater than a few
$10^{44}\,{\rm erg\,s}^{-1}$  indicate the presence of a
quasar. Therefore in our IR-bright galaxy sample, IRAS~08572+3915N and
Mrk~463E \citep[see also][]{Mazzarella1991}, which are not identified
as optical quasars, 
would be good candidates for buried quasars. Interestingly, these two are in close interacting systems and not in
fully merged ULIRGs, thus suggesting that quasars can be formed before
the end of the interaction process \citep[see e.g.][]{Hopkins2006, Hopkins2008, RamosAlmeida2011RG}.
However, as noted in Section~\ref{sec:deblendIRS}, for
the fully embedded nuclear source of IRAS~08572+3915N
the $12\,\mu$m AGN luminosity should be considered an upper limit.
Therefore, from now on we we will 
refer as the sample of IR-bright quasars to the following six objects:
IZw1, Mrk~1014, Mrk~231, IRAS~13349+2438, Mrk~463E, and Mrk~478.

The AGN fractional contributions within the slit (typical physical regions of a
few hundred parsecs)
at $12\,\mu$m and in the  $\sim 5-15\,\mu$m spectral range
for the IR-bright galaxy nuclei (see Table~\ref{tab:deblendIRS_ULIRG})
show a large variation going from
$0.4$ (NGC~6240N and IRAS~17208$-$0014) to nuclei completely dominated by the
AGN emission (e.g., Mrk~1014 and IRAS~13349+2438).  The median value
of the AGN fractional contribution at $12\,\mu$m
for the IR-bright galaxy nuclei is 0.78 (within the slit). 
This high value is not surprising because our sample selection required
  relatively compact emission and this tends to select IR-bright galaxies with
  AGN-dominated mid-IR  fluxes within the slit
  (see discussion in Section~\ref{sec:sample}). 
The nuclear (typical physical scales of hundreds of pc) mid-IR
emission of the majority of quasars in our sample appears to be
dominated by AGN emission (median
  value AGN contribution at $12\,\mu$m of 0.87), which is 
consistent with the AGN fractional contributions at
$24\,\mu$m derived from {\it Spitzer}/IRS spectroscopy by
\cite{Shi2007}.

\begin{figure*}

	\includegraphics[width=0.9\columnwidth]{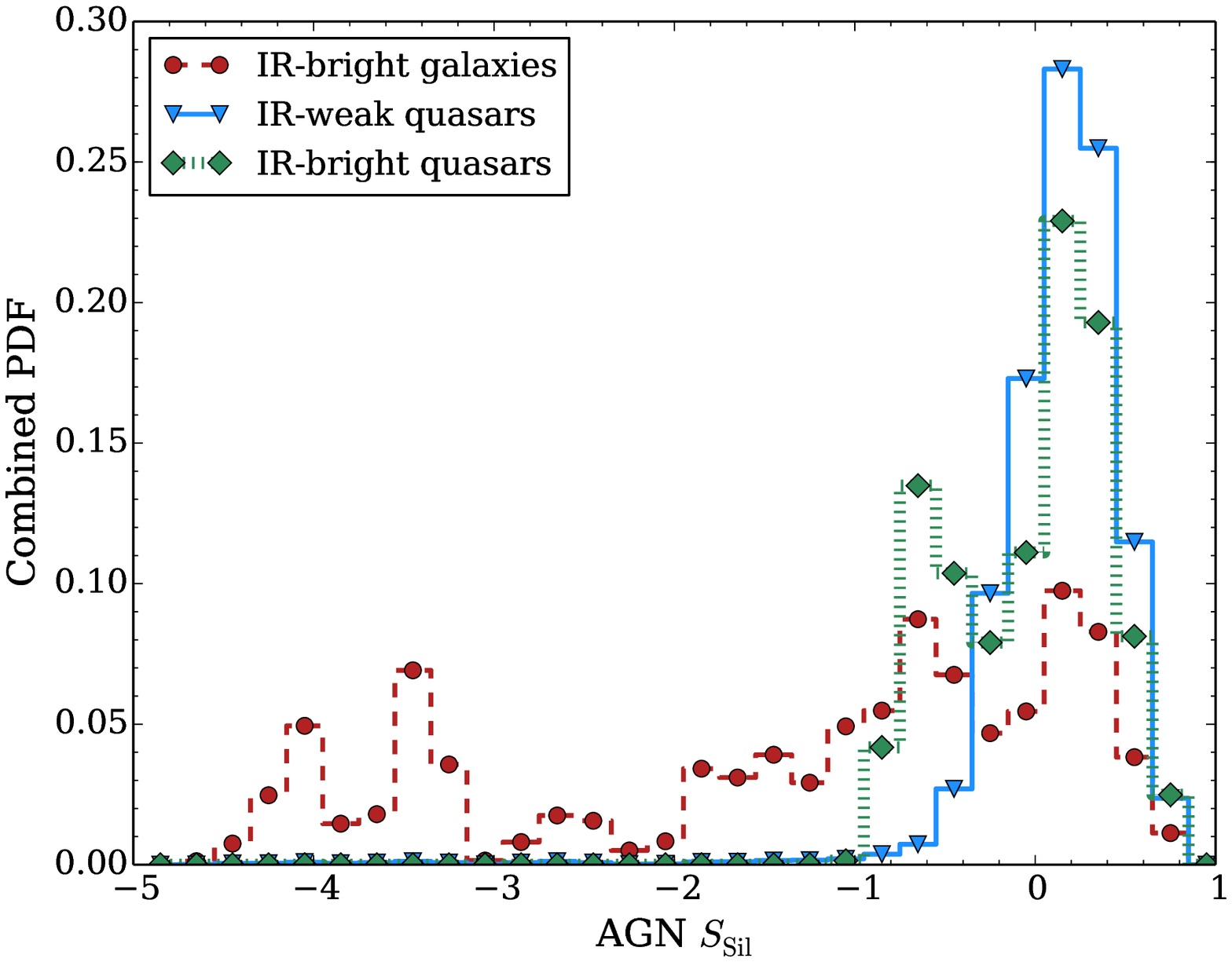}
	\includegraphics[width=0.9\columnwidth]{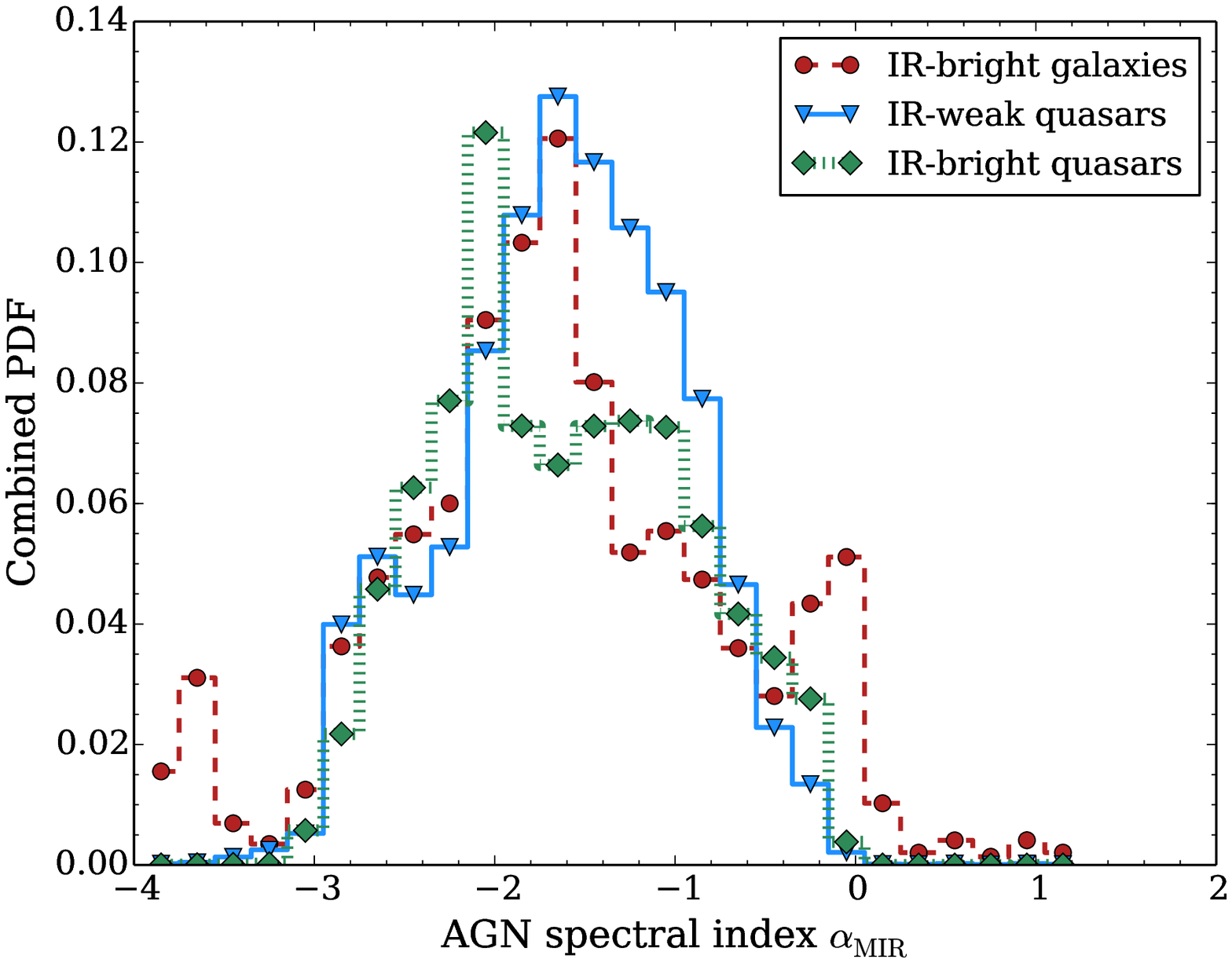}
    \caption{Combined probability distribution functions of the AGN
      mid-IR ($5-15\,\mu$m) spectral index (right panel) and strength of the
      silicate feature (left panel) derived with {\sc
        deblendIRS}. The IR-bright galaxy sample includes the IR-bright quasars
      and contains 16 nuclei, the IR-weak quasar sample contains 10 nuclei, and
      the IR-bright quasar sample contains 6 nuclei. } 
    \label{fig:combinedPDF}
\end{figure*}

 We can use the average
AGN and quasar templates of \cite{Mullaney2011}, which have $L^{\rm AGN}_{\rm IR}/\nu L_\nu^{\rm AGN}(12\mu{\rm m})=2.3$
and 1.6, respectively, to derive the IR ($8-1000\,\mu$m) 
emission of the AGN. We find that the IR emission of the AGN  
contribute between 1\% (NGC~6240N and IRAS~17208$-$0014) and $\sim
68\%$ (Mrk~463E) of  the total IR luminosity of the system in
our IR-bright galaxy sample  with a median value of  10\%. This is in contrast
with the AGN mid-IR dominance within the central hundred parsecs and
indicates that the IR-bright galaxies in our sample have extended star formation
over several kiloparsecs (see also Section~\ref{sec:SF}).  The largest AGN contributions among the
IR-bright galaxies are, not surprisingly, for the IR-bright quasars. These estimates are in good agreement
with the average $35-40\%$
AGN bolometric contribution estimated for local ULIRGs \citep{Veilleux2009ULIRG,Nardini2010} and $5-15\%$
for local LIRGs \citep{Petric2011,AAH2012}  applying different methods to the {\it Spitzer}/IRS 
spectra.
For the quasars the AGN contribution to the IR luminosity is higher
than for the IR-bright galaxies, with values in the range $30-70\%$
and a 
median contribution of 60\%. 

\subsection{AGN silicate feature and mid-IR spectral index}\label{sec:AGNalphaSSil}
For each  nucleus in the IR-bright galaxy and quasar samples, we list in
Tables~\ref{tab:deblendIRS_ULIRG} and \ref{tab:deblendIRS_QSO} respectively
the median values and $1\sigma$ confidence intervals of the
fitted strengths of the  silicate
feature and mid-IR spectral index of the AGN component. As can be seen from
these tables, the AGN hosted in IR-bright galaxies show a broader range of 
strengths of the silicate feature than the IR-weak quasars. The latter 
tend to have values consistent with the feature being slightly in emission,
as also found by other studies using {\it Spitzer}/IRS spectroscopy
\citep[see e.g.][]{Shi2006,Thompson2009}. In terms of the AGN spectral
index in the $5-15\,\mu$m range, the two samples show
values that are similar to values derived for Seyfert galaxies
although some for slightly different spectral ranges
\citep[see e.g.][]{Thompson2009,Hoenig2010,RamosAlmeida2011,HernanCaballero2015,AAH2016}.

To make a statistical comparison of the derived mid-IR AGN properties of
IR bright galaxies and quasars we combined the individual PDF of the
the mid-IR $5-15\,\mu$m spectral index and the strength of the silicate feature
from the {\sc deblendIRS} spectral
decomposition for the nuclei of the samples.
We show the resulting
combined PDFs in Fig.~\ref{fig:combinedPDF} and the statistics
in Table~\ref{tab:statisticsPDF}. 
We note that the relatively small size of our samples means that some of the
  observed peaks in the combined PDF are due to individual nuclei with relatively
  well constrained values of the AGN $S_{\rm Sil}$ and $\alpha_{\rm MIR}$ (see discussion
  below). 
  However, we prefer to show the combined PDFs rather than the PDF for the individual
  galaxies as the PDF for some individual objects are quite broad whereas the combined PDF
  contain more information. 

The expectation value for $\alpha_{\rm MIR}$ derived from the
combined PDFs (median of the distribution) is similar for the samples of IR-bright
galaxies and the IR-weak quasars
($\alpha_{\rm MIR} =-1.7\,{\rm to\,} -1.8$). However, the distribution for the 
nuclei of IR-bright galaxies shows a
slightly broader tail toward flatter mid-IR spectral  indices. On the
other hand, the distributions of the strengths  of the silicate
features for the AGN component are markedly different for the two samples. The IR-weak
quasars show a narrow distribution centered at $S_{\rm Sil}=0.07$ (the
silicate feature slightly in emission),
whereas the distribution for the AGN hosted by IR-bright galaxies
(including IR-bright quasars, but see below) is
much broader with a median value of 
$S_{\rm Sil}=-0.90$. These results are similar to the findings for
ULIRGs and  optical quasars by \cite{Veilleux2009ULIRG} based on the {\it
  Spitzer}/IRS spectroscopy.

\begin{figure}
\hspace{-0.5cm}
\resizebox{1.1\hsize}{!}{\rotatebox[]{0}{\includegraphics{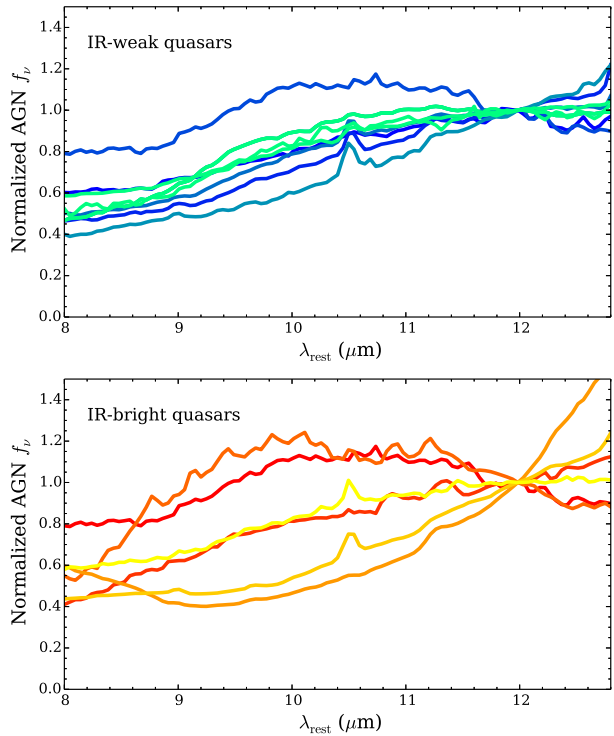}}}
\vspace{-0.5cm}
    \caption{The thick color lines are the {\sc deblendIRS} best-fit AGN components  of the ground-based mid-IR
    spectra of the IR-weak quasars (upper panel) and IR-bright
    quasars (lower panel), normalized at $12\,\mu$m.} 
    \label{fig:AGNbestspectra}
\end{figure}

We also produced the combined PDFs for the IR-bright quasars alone
(six nuclei excluding IRAS~08572+3915N), see bottom panels of
Fig.~\ref{fig:combinedPDF} and Table~\ref{tab:statisticsPDF}. The
best-fit AGN components for both IR-bright and IR-weak quasars
are relatively similar (see Fig.~\ref{fig:AGNbestspectra}), as also reflected by the median 
values of the combined PDFs of $S_{\rm Sil}$ and $\alpha_{\rm
  MIR}$. However, there are some slight differences in the 
$16\%-84\%$ percentiles of the combined PDF of $S_{\rm Sil}$.
There is a second peak in the distribution of
the AGN $S_{\rm Sil}$ of the IR-bright quasars at $S_{\rm Sil} \sim -0.8$, which is due to the
two "transitioning quasars" in our sample, namely, 
Mrk~463E and Mrk~231 (the two best-fit AGN components with the
$9.7\,\mu$m silicate feature in absorption in Fig.~\ref{fig:AGNbestspectra}). Conversely, the 
peak  at a value $S_{\rm Sil} \sim 0.1-0.2$ seen in the combined PDF of
the IR-bright galaxy nuclei is contributed by the IR-bright quasars (IZw1, Mrk~1014, IRAS~13349+2438, 
and Mrk~478), whereas the rest of the AGN components in the IR-bright
galaxy 
sample appear to be much more obscured.

\begin{table*}
	\centering
\caption{Statistics of the combined PDF for the fitted mid-IR AGN properties.}
\label{tab:statisticsPDF}
\begin{tabular}{@{}lcccccc@{}}
\hline
 
Sample & Number &  \multicolumn{2}{c}{AGN $S_{\rm Sil}$} & \multicolumn{2}{c}{AGN $\alpha_{\rm MIR}$} \\

 & & Median & [16\%, 84\%] & Median & [16\%, 84\%]\\
\hline
IR-bright galaxies & 16 & -0.90 & [-3.52, 0.09] &
 -1.79 & [-2.63, -0.73]\\
IR-weak quasars  & 10   & 0.07 & [-0.24, 0.33] & 
-1.68 & [-2.40, -1.04]\\
IR-bright quasars & 6 & -0.03 & [-0.68, 0.29] & 
-1.58 & [-2.39, -0.35]\\
\hline
\end{tabular}
\end{table*}

In Section~\ref{sec:evolution} we will investigate further the
possible  evolutionary connection between IR-bright galaxies and quasars and the
interaction stage of the
AGN mid-IR emission.

\subsection{Nuclear star formation activity}\label{sec:SF}
In this section we use the $11.3\,\mu$m PAH feature emission to study the
star formation activity in the nuclear regions of the IR-bright galaxy
sample. Of the 13 IR-bright nuclei where the ground-based spectral range allowed
measurements, we detected the $11.3\,\mu$m PAH feature (with a
detection significance better than $2\sigma$) in 7 of them (see
Table~\ref{tab:measurePAH}). The nuclear values of the  EW of 
the $11.3\,\mu$m PAH feature for most of the nuclei  in the IR-bright galaxy sample
 are below the typical values of
star forming galaxies \citep[$\sim 0.5-1\,\mu$m, see for instance][]{HernanCaballero2011}
because of the increased  AGN fractional contribution at $12\,\mu$m when using
high angular resolution data. For
the $11.3\,\mu$m PAH non-detections there still can be a small contribution from
star formation activity in the nuclear region. Our upper limits show
that sources with undetected $11.3\,\mu$m PAH features could still
have $11.3\,\mu$m PAH luminosities of a few $10^{41}\,{\rm erg \,s}^{-1}$.  
This shows
that detecting PAH emission may be difficult  when the AGN continuum dominates the
emission at $12\,\mu$m unless the spectrum has a high
S/N ratio. In galaxies with deep silicate 
features (e.g., IRAS~08572+3915N in our sample) it is possible that
the $11.3\,\mu$m PAH emission is obscured by the same foreground
absorber that obscures the AGN continuum 
\citep[see also discussion by][]{Veilleux2009ULIRG}. In fact for IRAS~08572+3915N, the 100\% AGN
contribution in the mid-IR is because this galaxy is fitted
with an ``obscured AGN'' template which  might be suffering from the same effect.

Since we also measured the $11.3\,\mu$m PAH feature in the {\it
  Spitzer}/IRS spectra, we computed the ratios between the nuclear and
the IRS $11.3\,\mu$m PAH luminosities (fifth column of
Table~\ref{tab:measurePAH}). The ratios for the IR-bright galaxy sample
indicate that  there is a large fraction with extended PAH emission from
the nuclear physical scales probed with the
ground-based data (hundreds of parsecs) to the few kpc scales probed
by the IRS slit. The two  nuclei in the IR-bright galaxy sample with 
the most extended $11.3\,\mu$m PAH emission (nuclear to IRS ratios of
0.08 and $<0.1$) are the two components of
Arp~299 (see Fig.~\ref{fig:Arp299}), as also revealed by the deep GTC/CanariCam imaging at
$8.7\,\mu$m with the Si-2 filter\footnote{This relatively narrow
  mid-IR filter probes the emission from the $8.6\,\mu$m PAH feature,
  see Figure~4 of \cite{DiazSantos2008}.}. The rest have ratios of between 0.1 and
0.5. For the nuclear spectra with only upper limits to the
$11.3\,\mu$m PAH feature ratios are consistent with compact emission
(ratios $0.5-1$).

For the 5 IR-weak quasars where the ground based spectra covered
the $11.3\,\mu$m PAH feature (see Table~\ref{tab:measurePAH}), we only detected the feature in Mrk~509
\citep[see also][]{Hoenig2010} with an EW of $0.011\pm
0.004\,\mu$m. This lack of detections is not surprising as the EW of
the $11.3\,\mu$m PAH feature measured from the {\it Spitzer}/IRS
spectra for the PG quasars in our sample are in the range
$0.068\pm 0.002\,\mu$m for Mrk~509 and $0.009 \pm0.002\,\mu$m for
Mrk~335. The $11.3\,\mu$m PAH feature of PG0804+761 remained
undetected even in the IRS spectrum. Since the ground-based slits are
about $\sim 7$ times narrower  than the IRS SL one,  higher AGN
contributions within the smaller slits would produce
even smaller EW of the PAH features. As can be seen from
Table~\ref{tab:measurePAH}, three quasars have formally  nuclear
upper limits to the  $11.3\,\mu$m PAH fluxes larger than the IRS
one due to the S/N ratio of the CanariCam
spectra. Clearly, the nuclear to IRS ratios have to be lower than one.

\begin{figure}
\hspace{-0.5cm}
\resizebox{1.12\hsize}{!}{\rotatebox[]{0}{\includegraphics{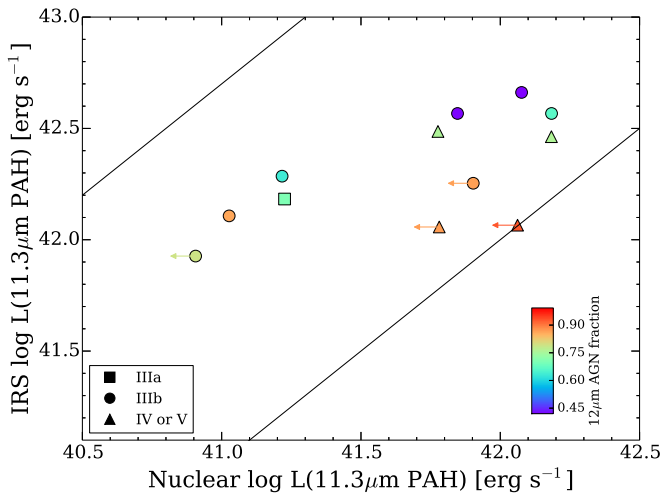}}}
    \caption{Comparison between the nuclear $11.3\,\mu$m PAH
      luminosities as measured from high angular resolution
      ground-based spectroscopy and kpc-scales $11.3\,\mu$m PAH
      luminosities as measured from {\it Spitzer}/IRS
      spectroscopy for the IR-bright galaxy sample. Symbols are color-coded
      by the AGN fractional contribution
      at $12\,\mu$m within the ground-based slits. Different
     symbols indicate the interaction class (see Table~1). The arrows
     indicate upper limits to the ground-based measurements of the PAH
     feature. The upper straight line
      indicates approximately uniformly distributed PAH emission whereas
      the lower line is the 1:1 relation, that is, all the emission is
    nuclear.} 
    \label{fig:SpitzervsCanariCamLPAH}
\end{figure}

In Fig.~\ref{fig:SpitzervsCanariCamLPAH} we plot the nuclear $11.3\,\mu$m PAH feature
luminosities
from ground-based data against those measured from the {\it
  Spitzer}/IRS spectra for the IR-bright galaxy sample. From the slit widths
and angular resolution we estimate that if the emission was uniformly
distributed we would expect a ratio between the two luminosities 
for the GTC/CanariCam spectra of approximately 50. In this figure, for
NGC~1614 we plot the nuclear (central $\sim 150\,$pc) $11.3\,\mu$m PAH luminosity estimated by
\cite{PereiraSantaella2015} rather than our $160 \,{\rm pc}\times
610\,{\rm pc}$ extraction, which
leads to a nuclear to IRS ratio of approximately 0.1. As can be seen from 
this figure, the $11.3\,\mu$m PAH emission of the majority of the
IR-bright galaxies while extended over several kpc is not uniformly
distributed. It does not appear to be very compact either
(approximately ratios of less than a few tenths), except perhaps for some of nuclei with nuclear upper
limits and NGC~6240S. 
 If the $11.3\,\mu$m PAH emission traces the recent
star formation activity in IR-bright galaxies, this indicates that star
formation is extended over several kiloparsecs and not
taking place uniformly distributed. Indeed, near-IR {\it HST} and
ground-based hydrogen recombination line observations show that in
local (U)LIRGs star formation takes place in high surface brightness H\,{\sc ii} regions 
\citep[see
e.g.][]{AAH2006, Arribas2012, PiquerasLopez2016}.
The morphologies traced by the GTC/CanariCam 
$8.7\,\mu$m imaging observations of those IR-bright galaxies in our sample with 
clearly resolved nuclear
regions \citep[see Figs.~\ref{fig:Arp299}  and
\ref{fig:NGC1614,NGC6240,IRAS17208}, and also][for other lower IR
luminosity LIRGs]{DiazSantos2008} also
confirm this.  

Our results agree with lower angular resolution {\it Spitzer}/IRS
studies that measured the spatial extent of 
mid-IR spectral features in a complete sample local LIRGs from the
Great Observatories All-sky LIRG survey \citep[GOALS, ][]{Armus2009}. 
While the extent of the  [Ne\,{\sc ii}]$12.8\,\mu$m
line emission is restricted to
the central few kpc of these galaxies, the PAH emission and in particular the
$11.3\,\mu$m feature is clearly more extended \citep{DiazSantos2010,
  DiazSantos2011}. Moreover, in (U)LIRGs with mid-IR AGN contributions $> 50$\%,
like many of those 
those analyzed in this work, the mid-IR continuum is more compact
than the PAH emission. This is especially prevalent in those systems experiencing the latest
stages of an interaction \citep{DiazSantos2011}. This also agrees
with our findings for some of the IR-bright galaxies in our sample studied in 
detail in \cite{AAH2014}. 

Using the calibrations of the SFR in terms of the luminosity of the
$11.3\,\mu$m PAH luminosity \citep[see][]{DiamondStanic2012, Shipley2016} we
find nuclear (not corrected for extinction) SFR\footnote{We note that \cite{DiamondStanic2012} when
  deriving their SFR recipe 
measured the fluxes of the $11.3\,\mu$m PAH features using {\sc
  pahfit} instead of a local continuum near the feature as in our work
(Section~\ref{sec:measurePAH}). The {\sc pahfit} fluxes are on average twice those
measured with a local continuum \citep[see][for more
details]{Smith2007}. We therefore multiplied our $11.3\,\mu$m PAH luminosities by
this factor when deriving the nuclear SFRs.}  for the 
nuclei of IR-bright galaxies (see Table~\ref{tab:measurePAH}) between $0.6\,M_\odot \,
{\rm yr}^{-1}$ for IC~694 and $7.6\,M_\odot \,
{\rm yr}^{-1}$ for NGC~6240S assuming an initial mass function
similar to that of \cite{Kroupa2002}. Taking into account the physical sizes
covered by the ground-based slits, these translate into nuclear SFR per unit area
between $18\,M_\odot \,
{\rm yr}^{-1}\, {\rm kpc}^{-2}$ for IRAS~17208$-$0014 and  $160\,M_\odot \,
{\rm yr}^{-1}\, {\rm kpc}^{-2}$ for NGC~6240S. These values are
typical of the nuclear values derived by \cite{PiquerasLopez2016} for
a sample of local (U)LIRGs based on extinction-corrected Pa$\alpha$ or Br$\gamma$ luminosities.

Taking into account the estimated nuclear SFR per unit area in our
sample of IR-bright galaxies
and the AGN contributions to the nuclear emission, we suggest that 
NGC~1614, IC~694 and NGC~6240S would be
potentially good
candidates for starburst-driven powerful outflows based on the
observed correlation between SFR per unit area and outflow velocity
\citep{Arribas2014, Heckman2015}. Indeed, \cite{Feruglio2013} and \cite{GarciaBurillo2015} have confirmed
the presence of starburst-driven outflows in the southern nucleus of
NGC~6240 and the nuclear region of NGC~1614, respectively, using
molecular gas data.

\begin{figure*}
\hspace{-0.5cm}
\resizebox{0.75\hsize}{!}{\rotatebox[]{0}{\includegraphics{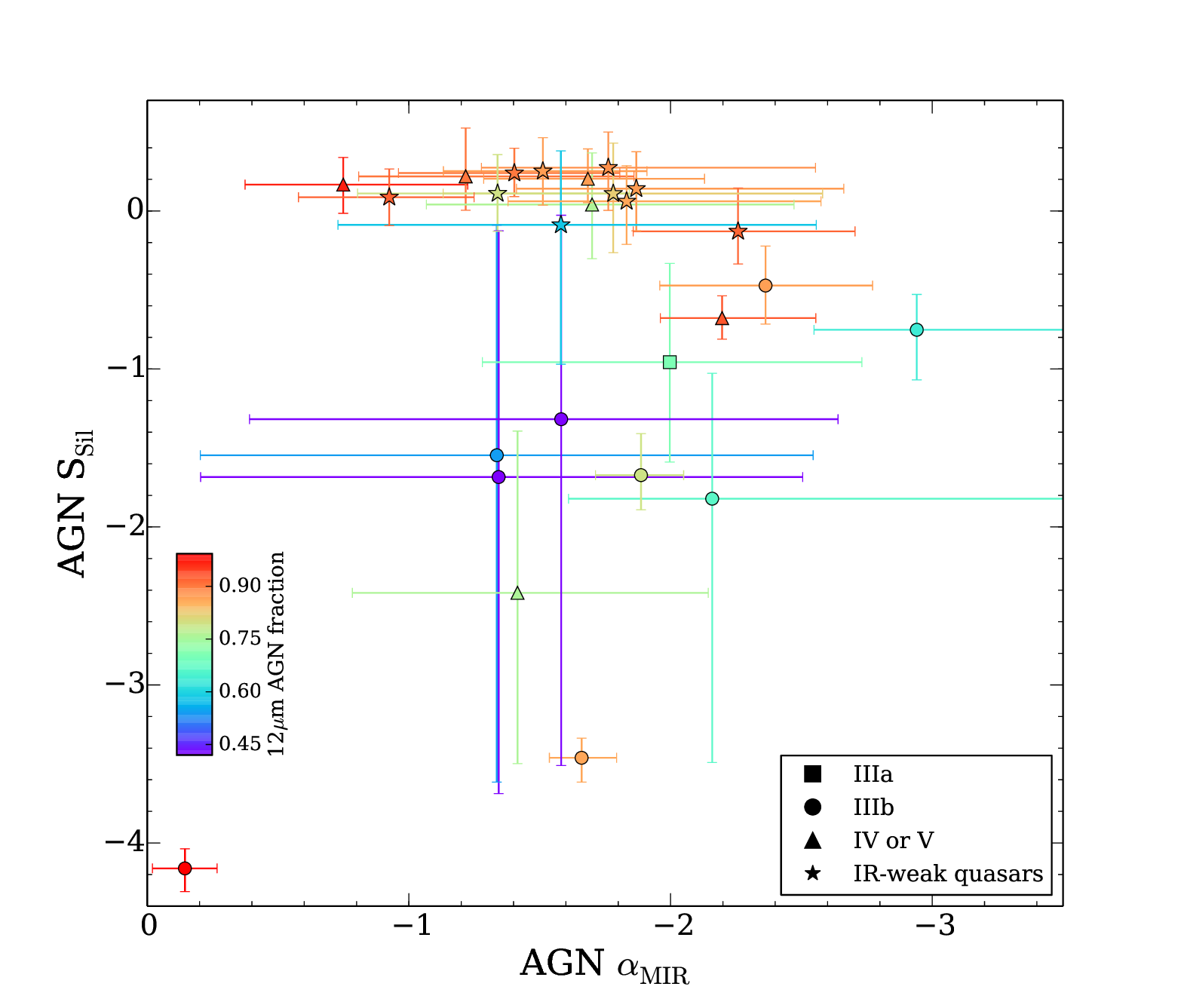}}}
    \caption{AGN mid-IR spectral index $\alpha_{\rm MIR}$ against the
      AGN strength of the
      silicate feature $S_{\rm Sil}$ as derived with {\sc deblendIRS} for both the
      sample of IR-bright galaxies and the sample of IR-weak quasars. 
     The error bars for the AGN mid-IR spectral index and strength of
     the silicate feature  are the $1\sigma$
    confidence levels (Tables~\ref{tab:deblendIRS_ULIRG} and \ref{tab:deblendIRS_QSO}) 
of the PDF of the fitted models. The
      measurements are color-coded by the AGN fractional contribution
      at $12\,\mu$m within the ground-based slits. The different
      symbols for the IR-bright galaxies indicate different interaction stages,
      whereas the IR-weak quasars are shown as star symbols. } 
    \label{fig:midIRalphavsSsil}
\end{figure*}

\section{Evolutionary sequence of mid-IR properties between
  IR-bright galaxies and quasars?}\label{sec:evolution}

\subsection{Nuclear Obscuring Material}\label{sec:obscuringmaterial}
In this section we look for evidence for an evolution of the obscuring
material around the AGN between the IR-bright
galaxies, IR-bright quasars and IR-weak quasars. To do so, we use the derived mid-IR AGN properties, namely,
the spectral index, the strength of
the silicate feature, and the fractional contribution to the mid-IR
emission. With regards to the AGN
properties, \cite{Haas2003} predicted an evolutionary change of the
mid-IR slope toward flatter indices from
``warm'' ULIRGs (those
dominated by AGN emission) to young quasars, evolved quasars
and finally dead quasars. ``Cool''  ULIRGs, even those hosting an AGN, are dominated by
star formation, and  emit most of their energy in the far-IR. In
``cool'' and ``warm'' ULIRG
systems the dust is distributed in a more spherical configuration 
and as the system evolves to a quasar the dust distribution is
predicted to develop a disk or torus-like configuration. 
From their figure~12 we can see that their predictions for the shape of
  the IR emission encompasses physical scales of less than 1\,kpc which compare well
  with the physical sizes probed by our ground-based spectroscopy.
Moreover, in the context of the predictions from clumpy torus models the mid-IR slope
and the strength of the silicate feature provide information about the
radial distribution of the clouds and the number of clouds along the
equatorial direction in the torus \citep{Hoenig2010, RamosAlmeida2014}.

\begin{figure*}

	\includegraphics[width=1.02\columnwidth]{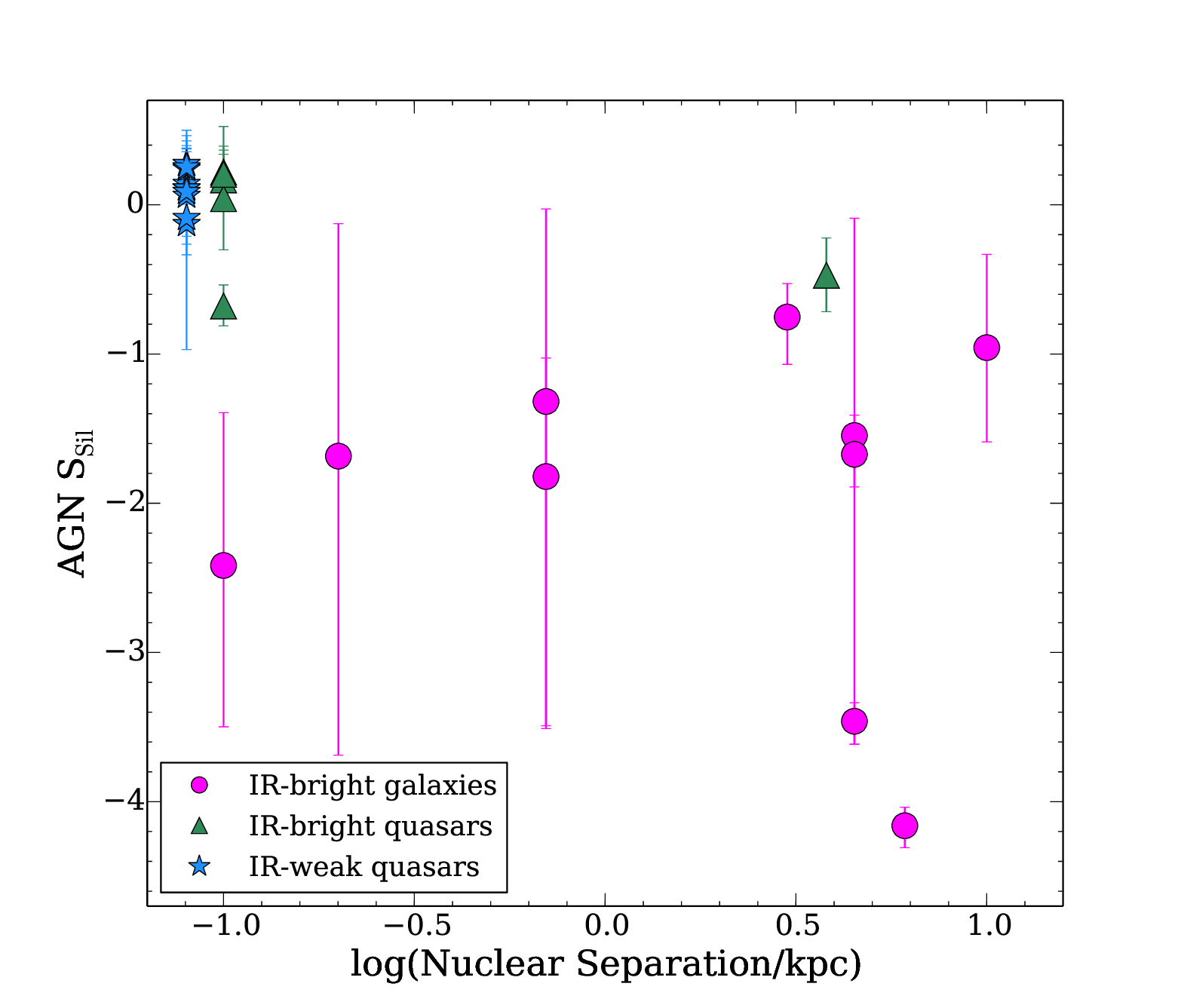}
	\includegraphics[width=1.02\columnwidth]{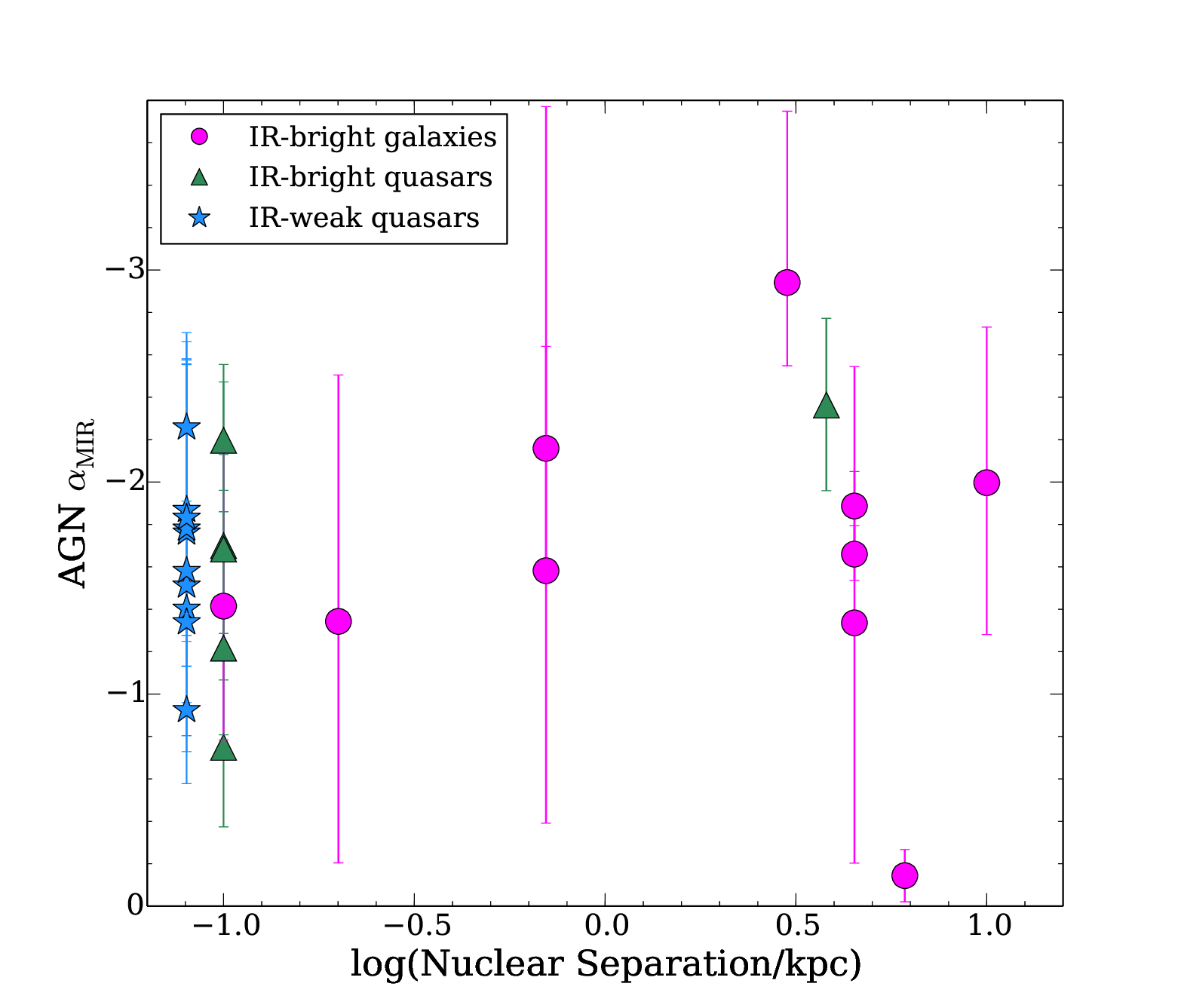}
	\includegraphics[width=1.02\columnwidth]{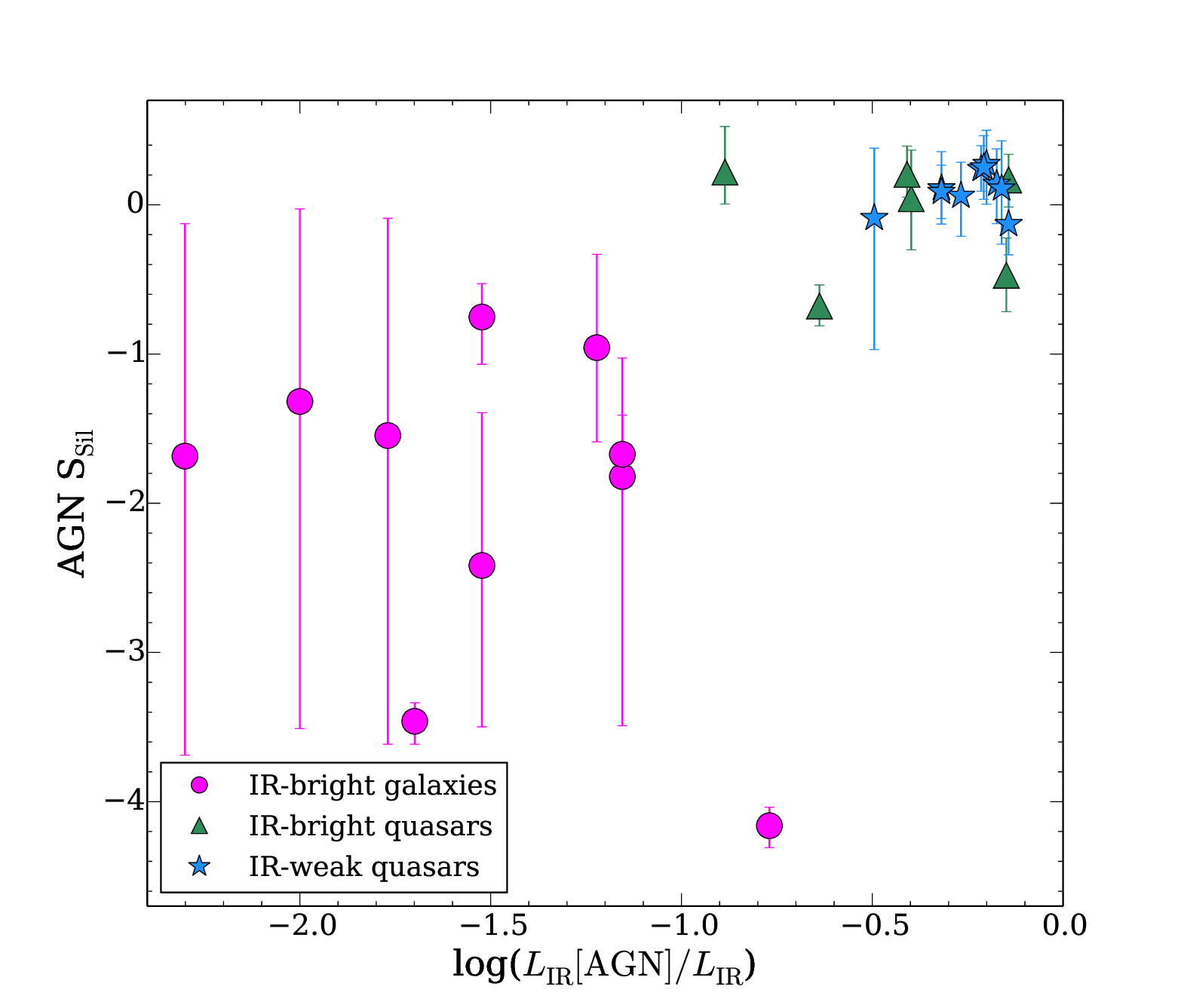}
	\includegraphics[width=1.02\columnwidth]{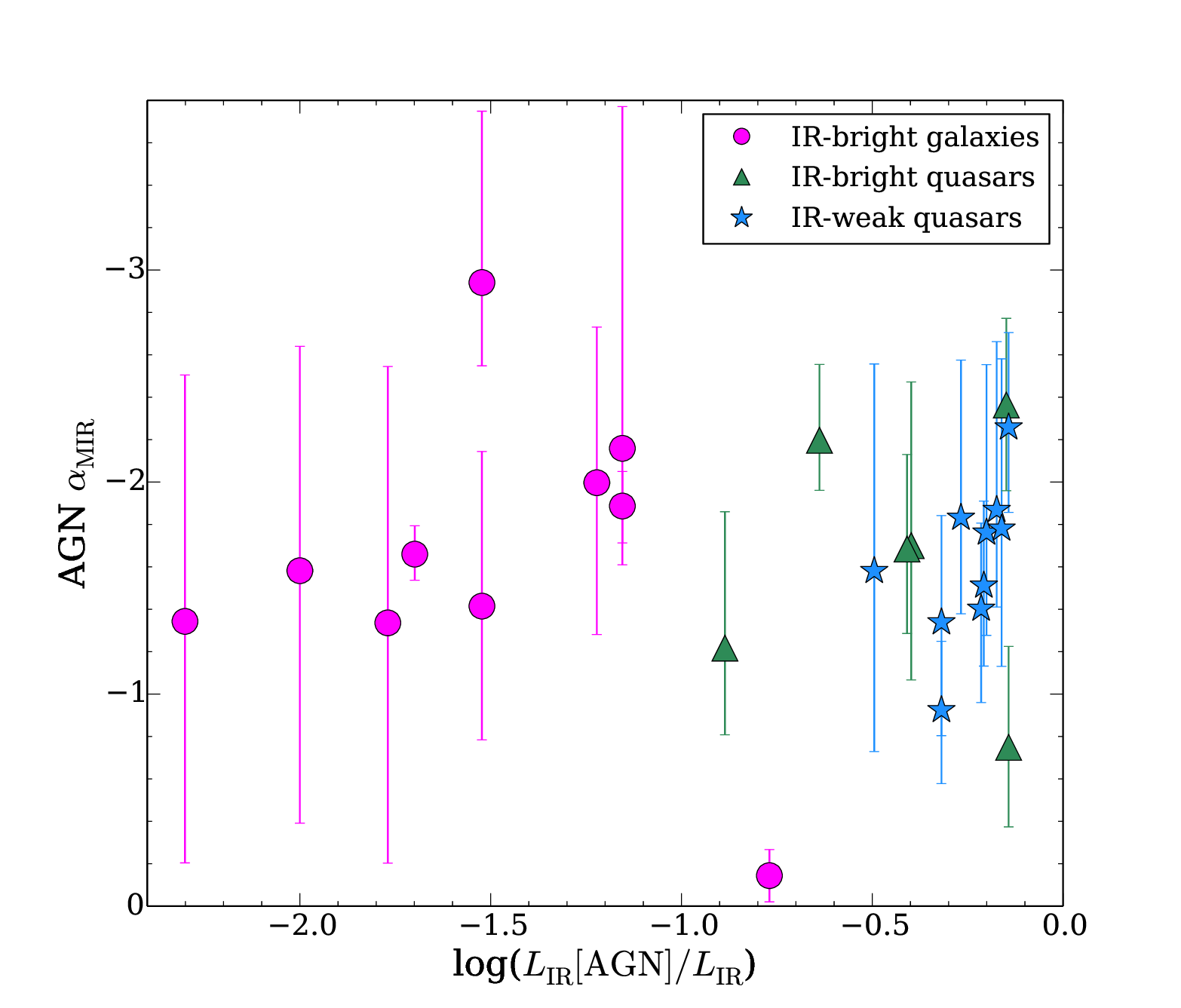}
    \caption{Comparison  between the derived
      AGN mid-IR spectral index $\alpha_{\rm MIR}$ and strength of
the silicate feature $S_{\rm Sil}$ and the projected nuclear
separation (upper panels) and the ratio between the AGN IR luminosity
and total IR luminosity of the system (bottom panels) for IR-bright
galaxies (only those not classified as quasars, 10  nuclei), IR-bright quasars and IR-weak quasars. The fully
merged IR-bright systems and IR-weak quasars are plotted at
projected nuclear separations of 0.1 and 0.08\,kpc for clarity.} 
    \label{fig:comparison_nucsep_IRratio}
\end{figure*}

Figure~\ref{fig:midIRalphavsSsil} compares the AGN mid-IR spectral
index to the AGN strength of the silicate feature
for the IR-bright galaxies and quasars as derived with {\sc deblendIRS}. The sources
are color-coded by the fractional AGN contribution at $12\,\mu$m and the IR-bright galaxies
are shown with their interaction class given in Table~\ref{tab:sampleULIRG}. 
Most of the IR-weak quasar 
sample and the IR-bright quasars (the latter recognised as having high AGN
fractional contributions and IV or V morphologies) are located in the region
around $S_{\rm Sil} =0$ (see also
Fig.~\ref{fig:comparison_nucsep_IRratio}) but span the same range of
mid-IR spectral indices as 
the rest of the IR-bright galaxy sample. This indicates that in general the nuclear
mid-IR emission of type 1 quasars can be explained with tori with a relatively
low number of clouds along the equatorial direction \citep[see][for
detailed modeling of the nuclear IR emission of PG quasars]{Mor2009,
  MartinezParedes2016}. 

As we also showed in Section~\ref{sec:AGNalphaSSil} for 
the combined PDF of $S_{\rm Sil}$ and $\alpha_{\rm MIR}$, the shapes of the AGN mid-IR emission of
 IR-bright  and IR-weak 
quasars do not differ significantly (see also Fig.\ref{fig:AGNbestspectra}). 
This seems to imply that the predicted evolution of the dust distribution
around these two types of quasars is taking place in short time
scales which cannot be resolved with our data. This agrees with the
observational result that red quasars make up only $15-20\%$ of the
quasar population \citep[see e.g.,][]{Glikman2012}.
Moreover, our classification between IR-bright and IR-weak
quasars is  based on the IR to
optical ratios which is likely  reflecting 
different levels of star formation activity in their host
galaxies as their far-IR emission is dominated by this process
\citep{Netzer2007}.  Finally, as discussed in
  Section~\ref{sec:sample}, our sample of quasars is likely mid-IR
  bright and thus it is possible we are missing more evolved quasars
  with different mid-IR
  spectral shapes.

  One  exception in this mid-IR view might be the transition quasar
Mrk~231 in our sample whose  AGN silicate feature is intermediate
between those of non-AGN dominated IR-bright galaxies and quasars 
suggesting the presence of extra dust components. In Mrk~231, the nuclear
silicate feature (measured on physical
scales of $<$400\,pc) would indicate an extinction of $A_V\sim 10\,$mag, using a foreground dust screen,
which is compatible with the value derived by  \cite{Veilleux2013} for the
obscuration of the broad absorption line region through which the
broad line region is observed. The AGN in the 
other two candidates to buried quasars (Mrk~463E and IRAS~08572+3915N) have strengths of the silicate
features which are not compatible with those measured in optically
identified quasars, with IRAS~08572+3915N showing in fact the most extreme
absorption feature in our sample of IR-bright galaxies and in general the local
(U)LIRG population \citep{Spoon2007}.  These three quasars could be
perhaps just before the
blowout phase of the quasar evolution \citep{Hopkins2006}.

As can be seen from Fig.~\ref{fig:midIRalphavsSsil}, a large fraction
of non-quasar IR-bright galaxies (8 out of 10) occupy a different
region from that of the quasars, especially in terms of the silicate
feature as we saw from the combined PDFs (see Fig.~\ref{fig:combinedPDF}).  
There is no clear trend of the AGN $5-15\,\mu$m spectral index with the AGN fractional
contribution for the IR-bright galaxy sample (see
Fig.~\ref{fig:comparison_nucsep_IRratio}, lower right panel). This would be expected if the AGN
fractional contribution 
were to give an indication of the evolutionary
stage of the AGN.  On the other hand, there is a  tendency for the AGN
component of  IR-bright galaxy nuclei with low ($<10\%$) AGN
fractional contributions to the total IR luminosity of the system to  show deeper silicate
features 
(Fig.~\ref{fig:comparison_nucsep_IRratio}, lower left panel). These have a mean
  value of $S_{\rm Sil}=-1.7$, whereas the nuclei with AGN fractional contributions above
  10\% have a mean value of $S_{\rm Sil}=-0.1$. This
is similar to the trend found by \cite{Veilleux2009ULIRG} for H\,{\sc ii}-like
optically classified ULIRGs to have deeper silicate features. As pointed out by
many works \citep{Levenson2007,
  GonzalezMartin2013, AAH2011, Hatziminaoglou2015, AAH2016}, deep
silicate features ($S_{\rm Sil}<-1$) cannot be reproduced by clumpy
torus models \citep[see e.g.][]{Nenkova2008,HoenigKishimoto2010},
indicating either extended dust 
components in the host galaxies \citep[see][]{Goulding2012} and/or deeply
embedded sources in more spherical configurations.
This is consistent
with \cite{Haas2003} evolutionary prediction of a more embedded  dust
distribution surrounding the AGN for the ``cool''  (i.e., star
formation dominated) ULIRG phase.

We do not find any clear
trends of the AGN mid-IR properties in terms of the interaction
stage (see Fig.~\ref{fig:midIRalphavsSsil}) or the projected nuclear separation (see the upper panels of
Fig.~\ref{fig:comparison_nucsep_IRratio}). 
For instance, the three AGN in the IR-bright galaxy sample with the deepest silicate
features are in interacting galaxies which are not fully merged (that
is classified as IIIb, with projected nuclear 
separations between the individual nuclei of less than
10\,kpc). However, the LIRG NGC~7469 (not plotted here), which is classified as IIIa (its
companion is IC~5283, located at a projected distance greater than 10\,kpc) has an AGN silicate
feature flat or slightly in emission \citep{Hoenig2010}, which is similar to the values
measured in the (fully merged) quasars in our sample and other Seyfert 1 nuclei
\citep{Thompson2009,Hoenig2010, AAH2016}. The other non-quasar IR
bright nuclei whose AGN is fitted with a deep
silicate feature is UGC~5101 which is believed to be fully merged. 

\begin{figure}
\hspace{-0.5cm}
\resizebox{1.15\hsize}{!}{\rotatebox[]{0}{\includegraphics{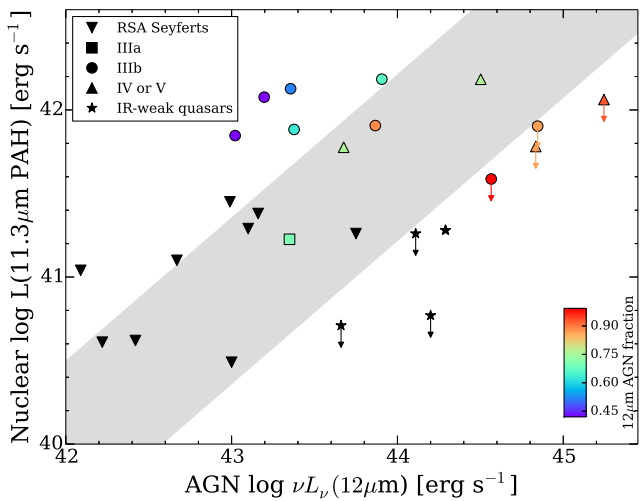}}}
    \caption{AGN rest-frame $12\,\mu$m  monochromatic luminosity against the
      nuclear $11.3\,\mu$m PAH  luminosity as proxies for the AGN luminosity
      and nuclear SFR, respectively, for the IR-bright galaxy and
      quasar samples. Symbols
      and colors as in Fig.~7. The $11.3\,\mu$m PAH luminosities and
      AGN $12\,\mu$m fractional contributions are
      from the ground-based mid-IR spectroscopy except for IC~694 and
      NGC~3690B (see text for more details). The filled star symbols
      are IR-weak quasars with detections or upper limits of the
      $11.3\,\mu$m PAH feature. The inverted triangles are RSA
      Seyferts for which the mid-IR spectroscopy probes physical
      scales of a few hundred parsecs. The shaded area indicates
        the locus of normal AGN, as defined by the location of the
        RSA Seyferts and assuming they follow a linear relation (see text for details).}
    \label{fig:LAGNvsLPAH}
\end{figure}

In summary, since the AGN mid-IR spectral index and the silicate feature
trace the properties of the obscuring material around the AGN
\citep[approximately the radial distribution of clouds and number of
clouds in the context of the clumpy dusty torus,
respectively, see][]{HoenigKishimoto2010} the only
clear trend is for both IR bright IR-weak quasars to 
show less obscuration than the rest of the  nuclei in
the IR-bright galaxy sample which tend to have lower AGN fractional
contributions. We
find no apparent relationship of these properties  with the interaction class of
the host galaxy or the projected nuclear separation between the nuclei. This confirms that there is not a single evolutionary
path for the local IR-bright galaxies into AGN dominated systems, as already discussed in previous works
\citep{Rigopoulou1999,Veilleux2009ULIRG,Farrah2009}. 

\subsection{Nuclear star formation vs. AGN luminosity}\label{sec:AGNvsSF}
In the last part of this work we investigate the relation between the
nuclear star formation activity and the AGN luminosity for our sample
of local IR-bright galaxies. Numerical simulations for major gas-rich mergers
\citep{Hopkins2008} predict a relation between these two quantities
with the correlation becoming more tightly coupled on smaller physical
scales \citep{HopkinsQuataert2010}. However, dynamical delays between
the peaks of star formation activity and
AGN activity are also predicted and are also a function of
the physical scales where the star formation is measured \citep{Hopkins2012}. From
the observational point of view this correlation has been observed for
local Seyferts in the Shapley-Ames (RSA) sample for nuclear SFR 
measured 
on circumnuclear  scales \citep[radius of $r=1\,$kpc, ][]{DiamondStanic2012} and 
on 100-pc nuclear scales
\citep{Esquej2014}.  Also \cite{Netzer2007} found a similar
correlation for PG quasars using the $7.7\,\mu$m PAH feature as measured
from {\it Spitzer}/IRS spectra and probing several kpc scales.

Figure~\ref{fig:LAGNvsLPAH} shows the AGN rest-frame $12\,\mu$m  monochromatic luminosity against the
      nuclear $11.3\,\mu$m PAH  luminosity as proxies for the AGN
      luminosity \citep[see][and references therein]{Asmus2015}
      and nuclear SFR, respectively, for our  IR-bright galaxy and quasar samples. Given
      the distances of the IR-bright galaxies and quasars in our sample the ground-based mid-IR
      spectroscopy probes physical scales of a few hundred
      parsecs (see Table~\ref{tab:measurePAH}). For the nearest LIRG in our sample, Arp~299, instead
      of the ground-based mid-IR PAH luminosities we show the {\it
        Spitzer}/IRS one which probes physical scales of
      approximately 700\,pc.

For a subset of Seyferts in the RSA sample, \cite{Esquej2014}  showed
that on nuclear scales (median of 60\,pc) there is a nearly linear
relation (with a 0.4\,dex scatter) between the SFR (calculated  from
the $11.3\,\mu$m PAH luminosity) and the black hole 
accretion rate (derived from the AGN bolometric luminosity). The RSA
Seyferts are drawn from a magnitude-limited catalog of galaxies and
considered to be one of the least biased samples of local AGN. As a
comparison we 
included in Fig.~\ref{fig:LAGNvsLPAH}  those RSA Seyferts for which, depending on
their distance, the ground-based or the {\it Spitzer}/IRS slits cover
physical regions of a few hundred parsecs, as in our samples. We note
that quasar Mrk~509 is also part of the RSA Seyfert sample. 

Since most of the RSA Seyferts are not currently experiencing
  major merger processes,
  we can use them
  to define the locus of {\it normal AGN} (in the sense of their activity
  being driven probably by
  more secular processes) in the plot showing the
  AGN luminosity (that is, black hole accretion rate) and the nuclear
  $11.3\,\mu$m PAH luminosity (nuclear star formation rate).  We plot this
  locus as a linear relation with a  $0.5\,$dex scatter, similar to the scatter
  measured on $r=1\,$kpc scales,
\citep[see][]{DiamondStanic2012}. As can be
seen from Fig.~\ref{fig:LAGNvsLPAH}, the quasars (both
IR-bright and IR-weak) lie at the high luminosity end of this
locus defined by the RSA Seyferts and are compatible with or lie just below the relation found for
lower luminosity AGN. This may suggest
that in some quasars the nuclear star formation activity is already
dimming as predicted by numerical simulations. Some of the other
IR-bright galaxy  nuclei (e.g., IC~694, NGC~1614, NGC~6240N,
IRAS~17208$-$0014) on the other hand, tend to show enhanced nuclear
star formation when compared to {\it normal} AGN. This suggests they are in an
earlier phase of the AGN evolution when star formation dominates the
energetics of the system. Note that \cite{AAH2013_LIRGS} reached a similar
conclusion for a sample of
local LIRGs at lower IR luminosities. However, this excess nuclear
star formation activity can
be observed for both
fully-merged IR-bright galaxies as well as individual nuclei of close
interacting systems.

\section{Summary and conclusions}\label{sec:conclusions}
The evolutionary connection between IR-bright galaxies
associated with gas-rich mergers and quasars was proposed nearly
thirty years ago.  We used ground-based mid-IR imaging and 
spectroscopy of a sample of 14 local (16 individual nuclei) 
IR-bright galaxies ($\log L_{\rm IR}\ge 11.4\,L_\odot$)
 and a comparison sample
of 10 optical quasars to investigate this connection. Among the
IR-bright galaxies, which are mostly in interacting or merger systems,
five are classified as IR-bright quasars based on
their IR luminosity to optical $B$-band ratios. 
We took advantage
of the high angular resolution (0.3-0.4\,arcsec) afforded by mid-IR
instruments on 8-10\,m-class telescopes to study nuclear physical
scales of hundreds of parsecs in both samples with an
improvement of almost a factor of 10 
with respect to {\it Spitzer}/IRS. This allows us to probe nuclear
scales where the black hole growth and star formation activity are
believed to be more tightly coupled.

Using the {\sc deblendIRS}
spectral decomposition tool we isolated 
the mid-IR emission due to dust heated by the AGN 
from that due to nuclear star formation for both samples. 
This allowed us to derive AGN mid-IR properties such as, the   $5-15\,\mu$m
spectral index ($\alpha_{\rm MIR}$), 
the strength of the $9.7\,\mu$m  silicate feature ($S_{\rm Sil}$), and
the AGN rest-frame $12\,\mu$m monochromatic 
luminosities, which provide information about the nuclear dust
distribution and the AGN contribution to the IR
luminosity. We also measured the $11.3\,\mu$m PAH feature emission on
nuclear and kiloparsec scales, the latter from {\it Spitzer}/IRS
spectroscopy, as proxies for the star formation activity in both
IR-bright galaxies and quasars. 

Our main results are as follows,
\begin{itemize}
\item
IR-bright galaxies and quasars show similar high AGN mid-IR 
contributions (median $\sim 80-90\%$) within the slits. However,  the AGN IR contribution to the 
total IR luminosity of IR-bright galaxies is significantly lower
(between 1\% and 70\%, median 10\%) than in
quasars (median 60\%)  indicating that the former have significant contributions from
star formation activity. 

\item 
The shapes of the AGN mid-IR emission of  IR-bright and IR-faint 
quasars do not differ significantly, as demonstrated by
the similarity of the combined PDF of the derived 
AGN $5-15\,\mu$m spectral indices (medians $\alpha_{\rm MIR} = -1.6$
and $\alpha_{\rm MIR} = -1.7$)
and 
strength of the $9.7\,\mu$m silicate feature (medians $S_{\rm
  Sil}=-0.03$, that is, feature slightly in emission and $S_{\rm
  Sil}=0.07$, feature slightly in absorption).
This seems to indicate that the predicted evolution of the nuclear dust distribution
around these two types of quasars is taking place in short time
scales and the differences in terms of IR to $B$-band ratios are likely due
to different levels of star formation activity in their host
galaxies.  However, our sample of quasars is likely to be mid-IR
  bright and therefore miss some of the more evolved quasars.

\item 
Based on the observed nuclear silicate features 
IR bright and IR-weak quasars are likely to contain less nuclear obscuring material than the rest of the  nuclei in
the IR-bright galaxy sample (median $S_{\rm Sil}=-0.90$). For the IR-bright galaxies we found no
clear trend of the AGN mid-IR properties  ($S_{\rm Sil}$ and
$\alpha_{\rm MIR}$) with the interaction class of
the host galaxy or the projected nuclear separation. This confirms that there is not a single evolutionary
path for the evolution of the dust distribution around  the AGN  hosted by  local
IR-bright galaxies. 

\item 
From the comparison of the $11.3\,\mu$m PAH fluxes between  nuclear
scales (ground-based spectroscopy) and kiloparsec scales ({\it Spitzer}/IRS
spectroscopy),  we concluded that   star formation 
in many local IR-bright galaxies is extended over several kiloparsecs
and not uniformly distributed but rather taking place in individual bright star forming
regions. The nuclear SFR of the IR-bright galaxies in our sample range 
between $0.6\,M_\odot \,
{\rm yr}^{-1}$ for IC~694 and $7.6\,M_\odot \,
{\rm yr}^{-1}$ for NGC~6240S. For the physical sizes
covered by the ground-based slits, these translate into nuclear SFR per unit area
between $18\,M_\odot \,
{\rm yr}^{-1}\, {\rm kpc}^{-2}$ for IRAS~17208$-$0014 and  $160\,M_\odot \,
{\rm yr}^{-1}\, {\rm kpc}^{-2}$ for NGC~6240S.

\item
  IR-bright and IR-weak quasars have more luminous AGN
    (using the AGN $12\,\mu$m luminosity
as a proxy) and higher nuclear
    (a few hundred parsecs) SFR (using the
$11.3\,\mu$m PAH luminosity as a proxy) than local
Seyferts. For their luminosities and compared to local Seyferts,
the nuclear star formation activity of some local quasars is already 
dimming, as predicted by numerical simulations of major mergers. Some of the
other IR-bright galaxy  nuclei, however, show enhanced nuclear
star formation at a given AGN $\nu L_\nu(12\mu{\rm m})$  when compared
to local Seyferts, indicating that they are in an
earlier star-formation dominated phase of the interaction process.
However, this excess nuclear star formation activity is
observed in  both
fully-merged IR-bright galaxies as well as individual nuclei of close
interacting systems. 
\end{itemize}

The results of this paper highlight the evolutionary complexity of the
AGN and star formation energetics in the the nuclear regions of local
IR-bright galaxies and quasars. The  integral field unit of the mid-IR MIRI
instrument \citep{Rieke2015, Wright2015} on the James Webb Space
Telescope will 
produce sensitive observations with similar angular resolutions to the
ground-based data analyzed here while providing a broad spectral range
($5-28.5\,\mu$m) and high spectral resolution ($R \sim 1500-3500$). This will
allow us to observe direct indicators of the presence of an AGN (i.e.,
detection of 
high excitation emission lines), probe the nuclear star formation
activity with different tracers, and obtain the kinematics of the nuclear regions of local
IR-bright galaxies and quasars.

\section*{Acknowledgements}

We thank an anonymous referee for valuable comments that helped improve
the paper.
We thank Santiago Garc\'{\i}a-Burillo for insightful discussions and
Sebastian H\"onig and Leonard Burtscher for making their fully reduced
spectra available to us. 
We acknowledge financial support from the Spanish Ministry
of Economy and Competitiveness through the Plan Nacional de 
Astronom\'{\i}a y Astrof\'{\i}sica grants AYA2012-31447
(A.A.-H. and A.H.-C.), and AYA2015-64346-C2-1-P (A.A.-H.),
which were partly funded by the FEDER programme, and 
AYA2015-70815-ERC and AYA2012-31277 (A.H.-C.) and AYA2012-32295
(M.P.-S. and L.C.). A.A.-H. is also partly
funded by CSIC/PIE grant 201650E036. R.P. acknowledges the
Oxford University Summer Research Programme. I.A. and M.M.-P. are partly
supported by Mexican CONACyT through research grant CB-2011-01-167291.
M.M.-P. acknowledges support by the CONACyT PhD fellowship programme.
C.R.A. acknowledges the Ram\'on y Cajal Program of the Spanish Ministry of
Economy and Competitiveness.
N.A.L. is supported by the Gemini Observatory, which is operated by
the Association of Universities for Research in Astronomy, Inc., on
behalf of the international Gemini partnership of Argentina, Brazil,
Canada, Chile, and the United States of America. T..D-S. acknowledges
support from ALMA-CONICYT project 31130005 and FONDECYT 1151239."

Based on observations made with the GTC, installed in the Spanish
Observatorio del Roque de los Muchachos of the Instituto de
Astrof\'{\i}sica de Canarias, in the island of La Palma. This research
has made use of the NASA/IPAC Extragalactic Database (NED)
which is operated by JPL, Caltech, under contract with the National
Aeronautics and Space Administration. This work is based in part on
observations made with the Spitzer Space Telescope, which is operated
by the Jet Propulsion Laboratory, California Institute of Technology
under a contract with NASA.






\appendix

\section{Examples of spectral decomposition with {\sc deblendIRS}}\label{appendix}

Even though {\sc deblendIRS} was initially designed to perform the
spectral decomposition of {\it
  Spitzer}/IRS spectra, it can also be used with ground-based data
covering the approximate spectral range $7.5-13\,\mu$m.  We
only needed to change 
the spectral range in the {\sc deblendIRS} configuration file to
reflect that covered by 
the ground-based
data. Additionally, from the list of AGN templates we removed those in
our sample selected as such \citep[see ][for more details]{HernanCaballero2015}.
We note, however, that the mid-IR fractional contributions of the STR,
PAH, and AGN components and the mid-IR spectral index $\alpha_{\rm
  MIR}$ are for the
$5-15\,\mu$m spectral range.

\begin{figure*}
	\includegraphics[width=0.9\columnwidth]{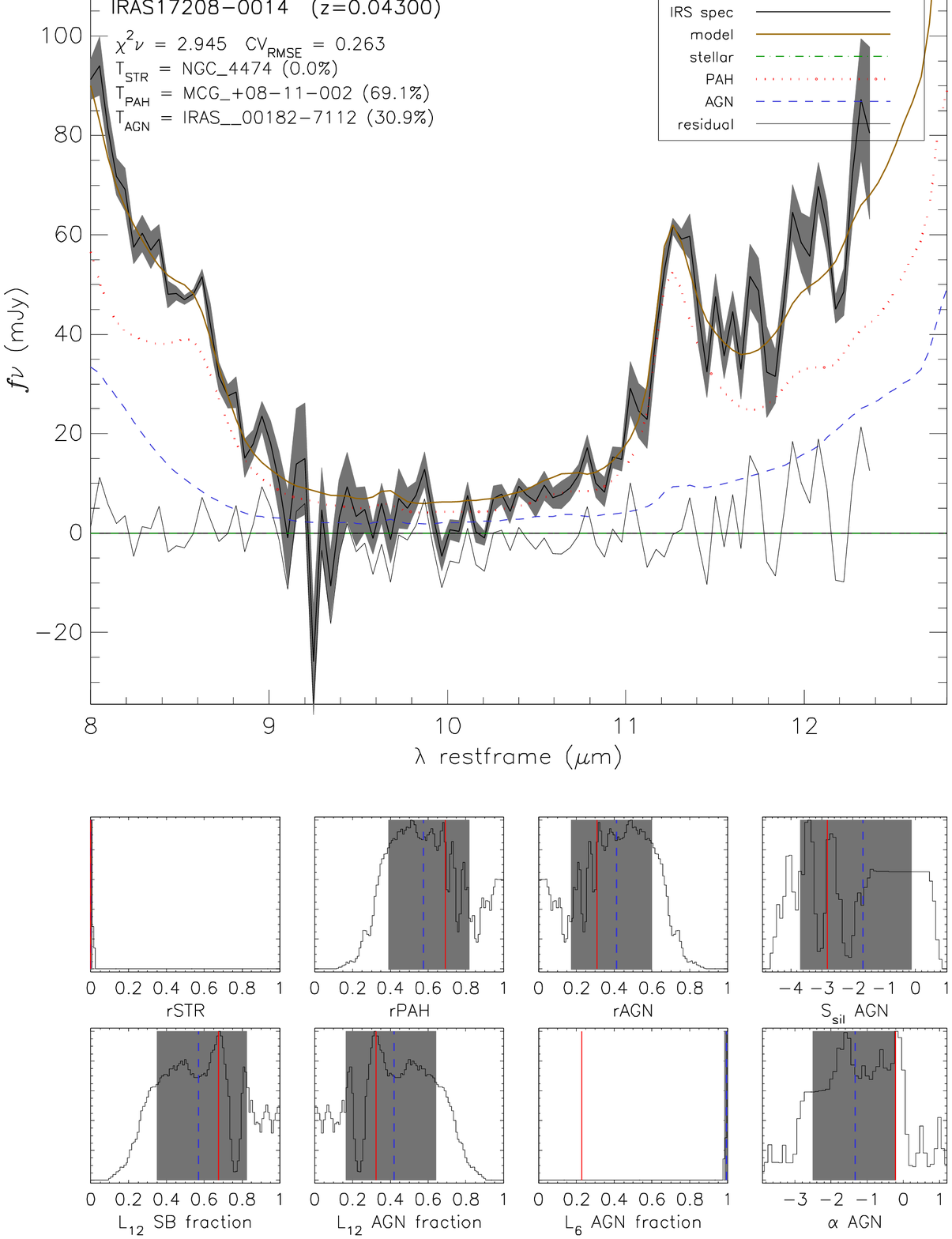}
	\includegraphics[width=0.9\columnwidth]{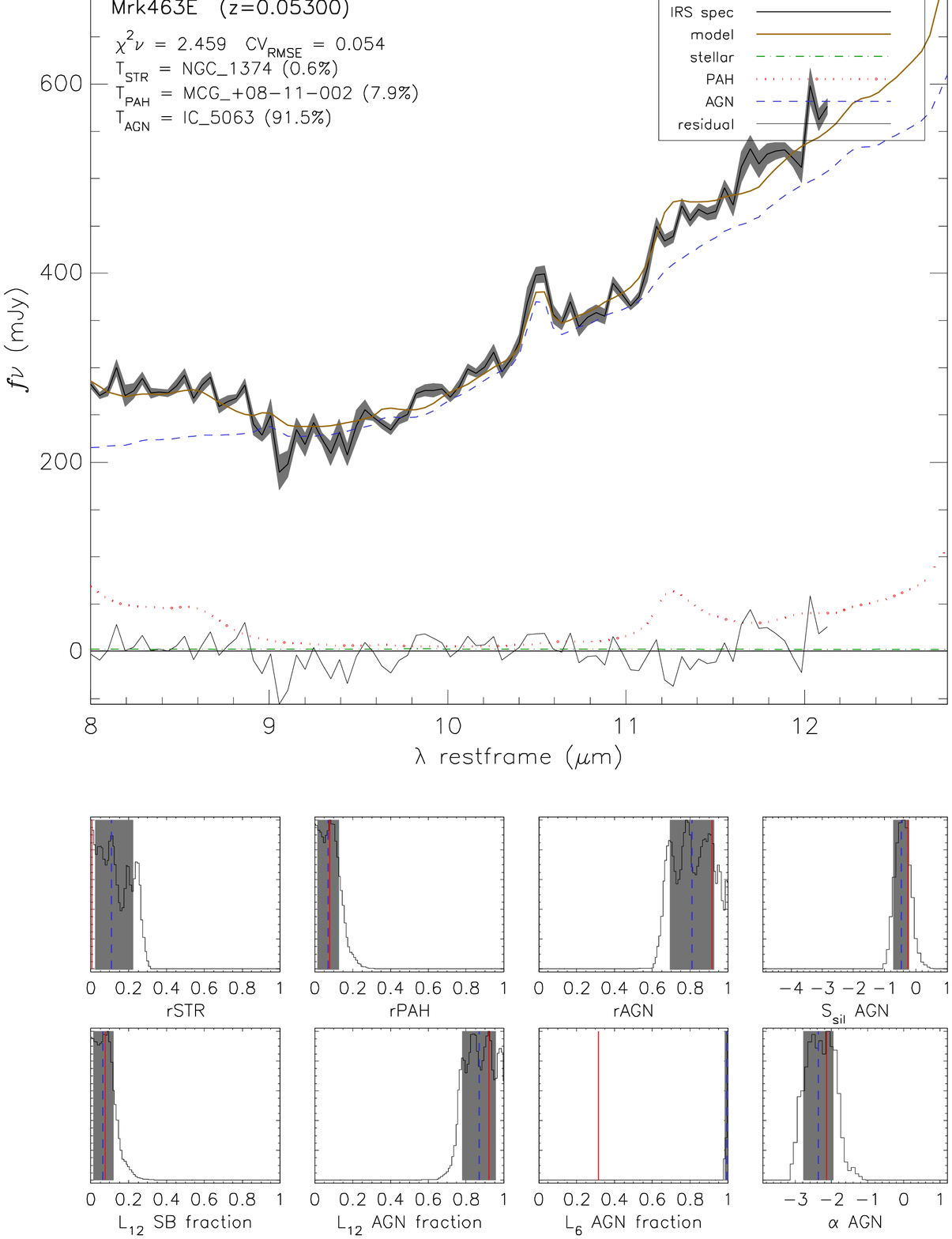}
        \caption{Examples of the graphical output of {\sc deblendIRS}
          for the spectral decomposition of the GTC/CanariCam 
          $8-13\,\mu$m nuclear spectra of an AGN-dominated IR-bright galaxy: Mrk~463E
          (right panel) and a star formation dominated IR-bright galaxy
          IRAS~17208$-$0014 (left panel). The top panels show the
          GTC/CanariCam spectrum in rest-frame (note that the label
          says ``IRS spec''), together with the
          best-fitting model (orange) and the stellar, PAH and AGN
          components in green, red, and blue.            The PDF in
          the lower panels are for the STR, PAH and AGN emission
          fraction (within the slit) in the $5-15\,\mu$m spectral
          range  rSTR, rPAH, and rAGN, respectively; the strength of
          the AGN silicate feature $S_{\rm Sil}$; the starburst and
          AGN fractional contribution  at $12\,\mu$m (within the slit)
          $L_{\rm 12}$ SB and $L_{\rm 12}$ AGN, respectively; the AGN
          fractional contribution at $6\,\mu$m  but since the
          rest-frame $6\,\mu$m is not covered by the CanariCam spectra
          the PDF of $L_6$ AGN fraction is meaningless; the AGN
          $\alpha_{\rm MIR}$ spectral index derived for the
          $5-15\,\mu$m 
          spectral range. For all the PDF the shaded regions represent
          the 1$\sigma$ confidence interval, that is, the 16\% and
          84\%
          percentiles, whereas the solid and dashed
          lines are the expectation value and the best-fit model value of the
          distributions, respectively.}  
    \label{fig:examplesdeblendIRS}
\end{figure*}

We show in Fig.~\ref{fig:examplesdeblendIRS} examples of the {\sc deblendIRS} graphical output of the spectral
decomposition of the GTC/CanariCam spectra of two IR-bright galaxy
nuclei in our sample, one dominated by the AGN emission (Mrk~463E) and
one dominated by star-formation (IRAS~17208$-$0014).

\begin{figure}
	\includegraphics[width=0.96\columnwidth]{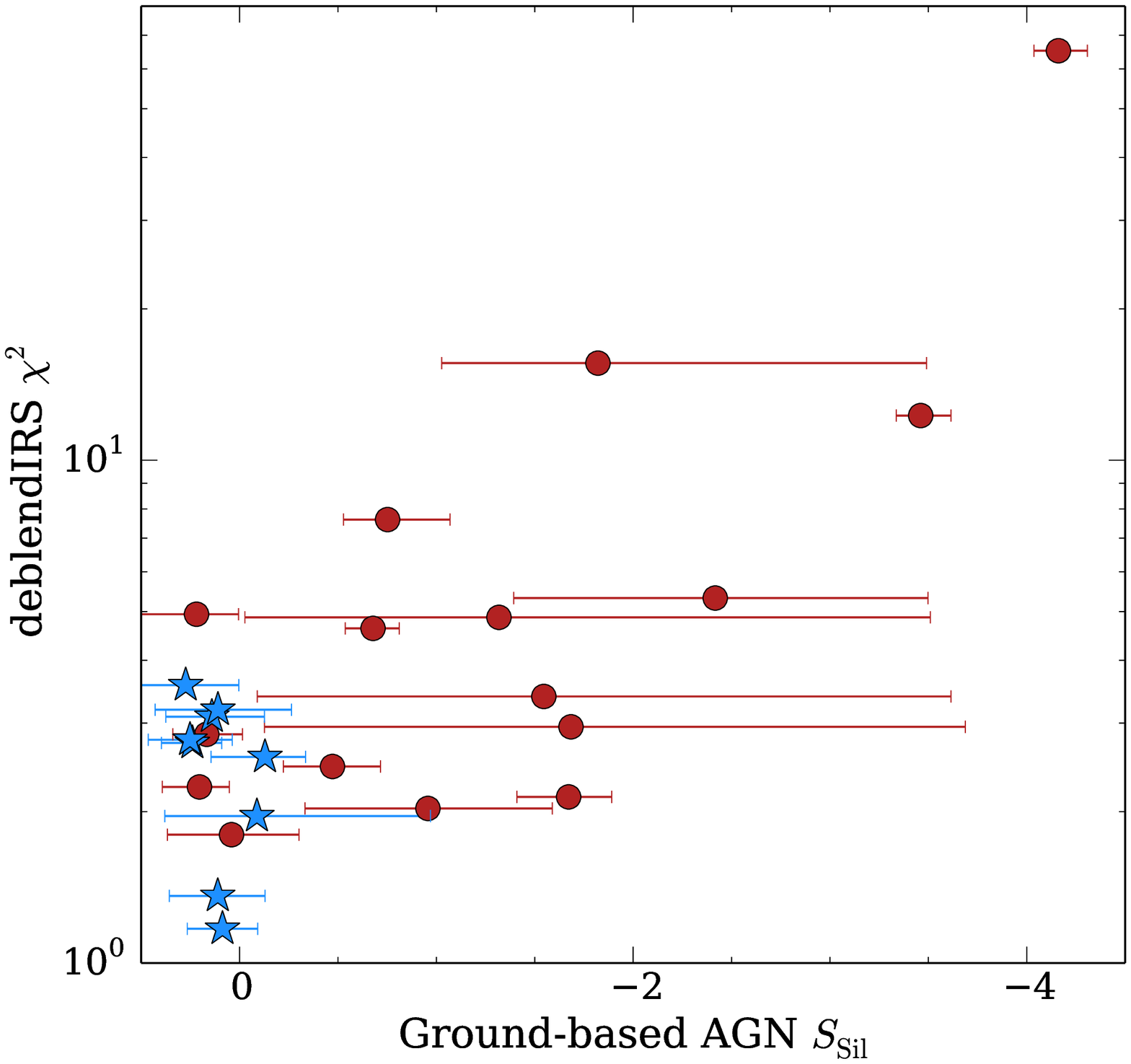}
        \caption{Comparison between the best-fit strength of the silicate feature of the
          AGN component plotted with
          the associated $1\sigma$ uncertainties and the 
          reduced $\chi^2$ value of the best fit obtained
          with {\sc deblendIRS} for the ground-based data. The  star symbols are the IR-weak
        quasars, whereas the circles are the IR-bright galaxies
        (including the IR-bright quasars).}  
    \label{fig:comparisonchisquareSsil}
\end{figure}

In Section~\ref{sec:deblendIRS} we remarked that based on the reduced $\chi^2$
values, the majority of the ground-based spectra of the IR-bright galaxies
and the quasars were well fitted with {\sc deblendIRS} using IRS templates.  However,
there are a few cases with relatively high $\chi^2$ values so we need
to determine 
if this is due to the lack of appropriate spectral templates in the {\sc deblendIRS}
database to fit deep silicate
features. In Fig.~\ref{fig:comparisonchisquareSsil}
we plot the reduced $\chi^2$ value of the best fit against the derived
strength of the silicate feature of the AGN component. As can be seen from this
figure, there is a tendency for worsening $\chi^2$  for deeper
silicate features cod nuclei with $\chi^2 > 10$ having the deepest silicate feature ($S_{\rm Sil}<-2$). This
might be in part due to the limited number of ``obscured AGN'' templates (22)
in the {\sc deblendIRS} database versus the rest of ``bonafide AGN'' templates
\citep[159, see][for further details]{HernanCaballero2015}. Also, the $\chi^2$ value
does not only depend on the appropriateness of the templates but
also the S/N ratio of the spectra.

\begin{figure*}
	\includegraphics[width=0.9\columnwidth]{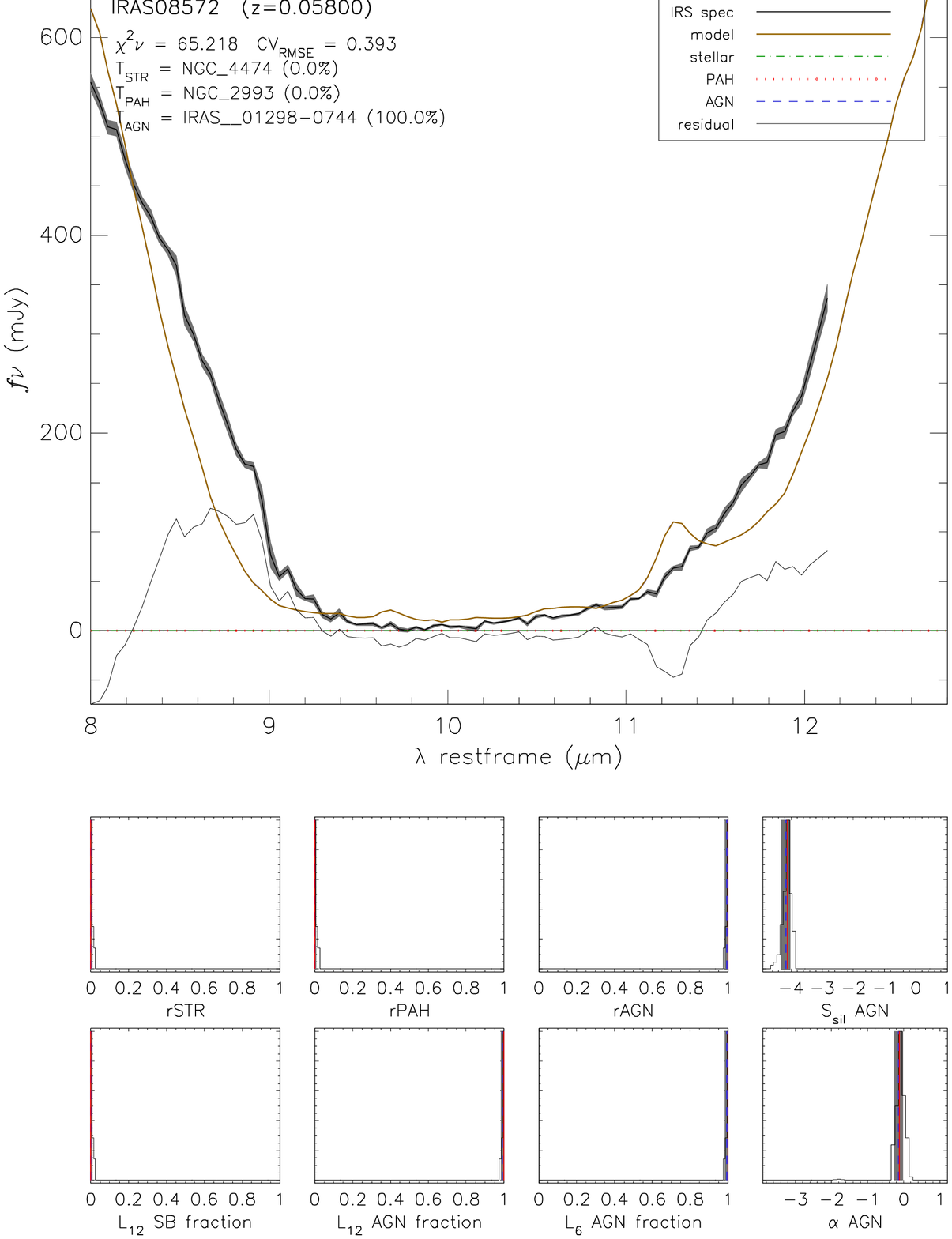}
	\includegraphics[width=0.9\columnwidth]{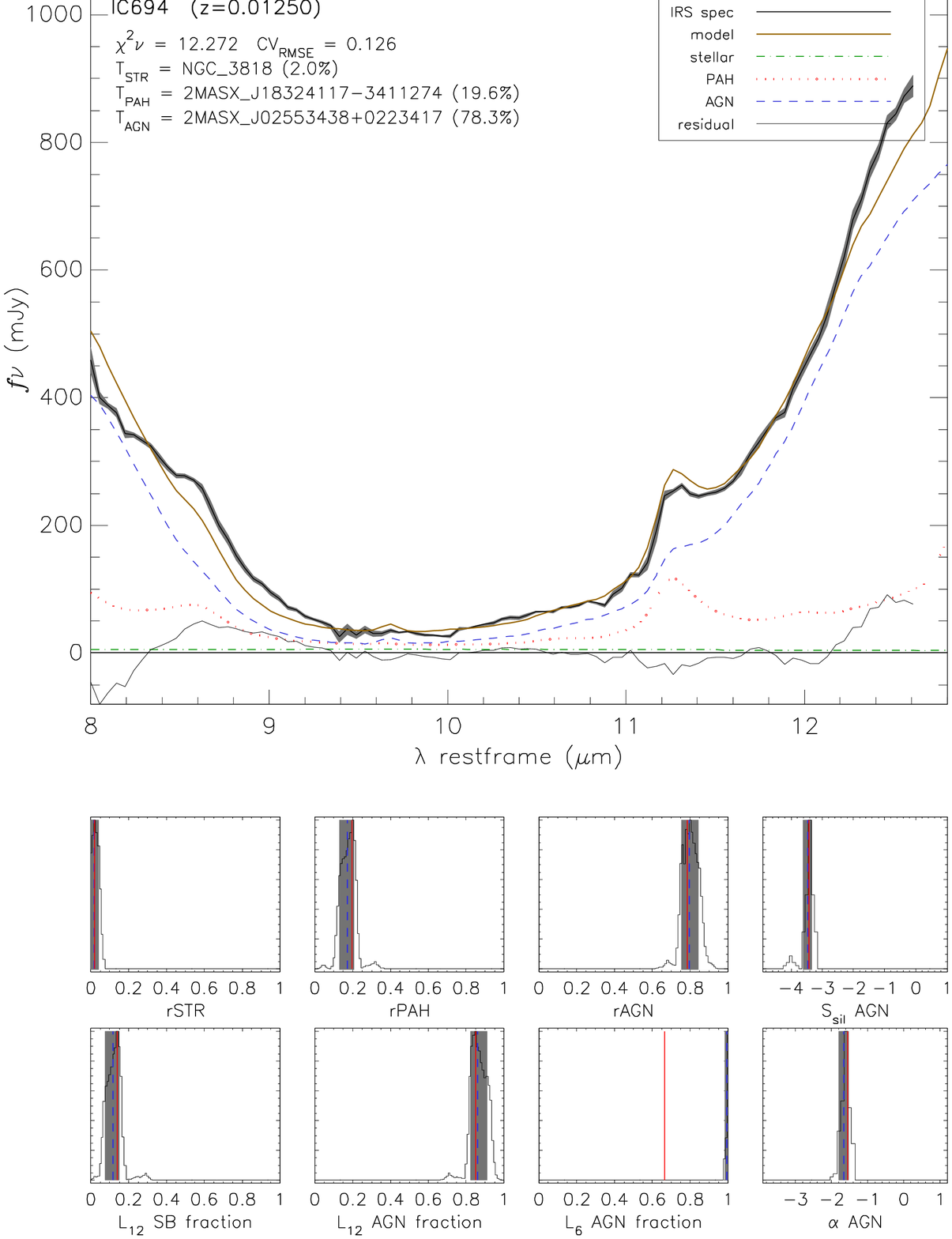}
	\includegraphics[width=0.9\columnwidth]{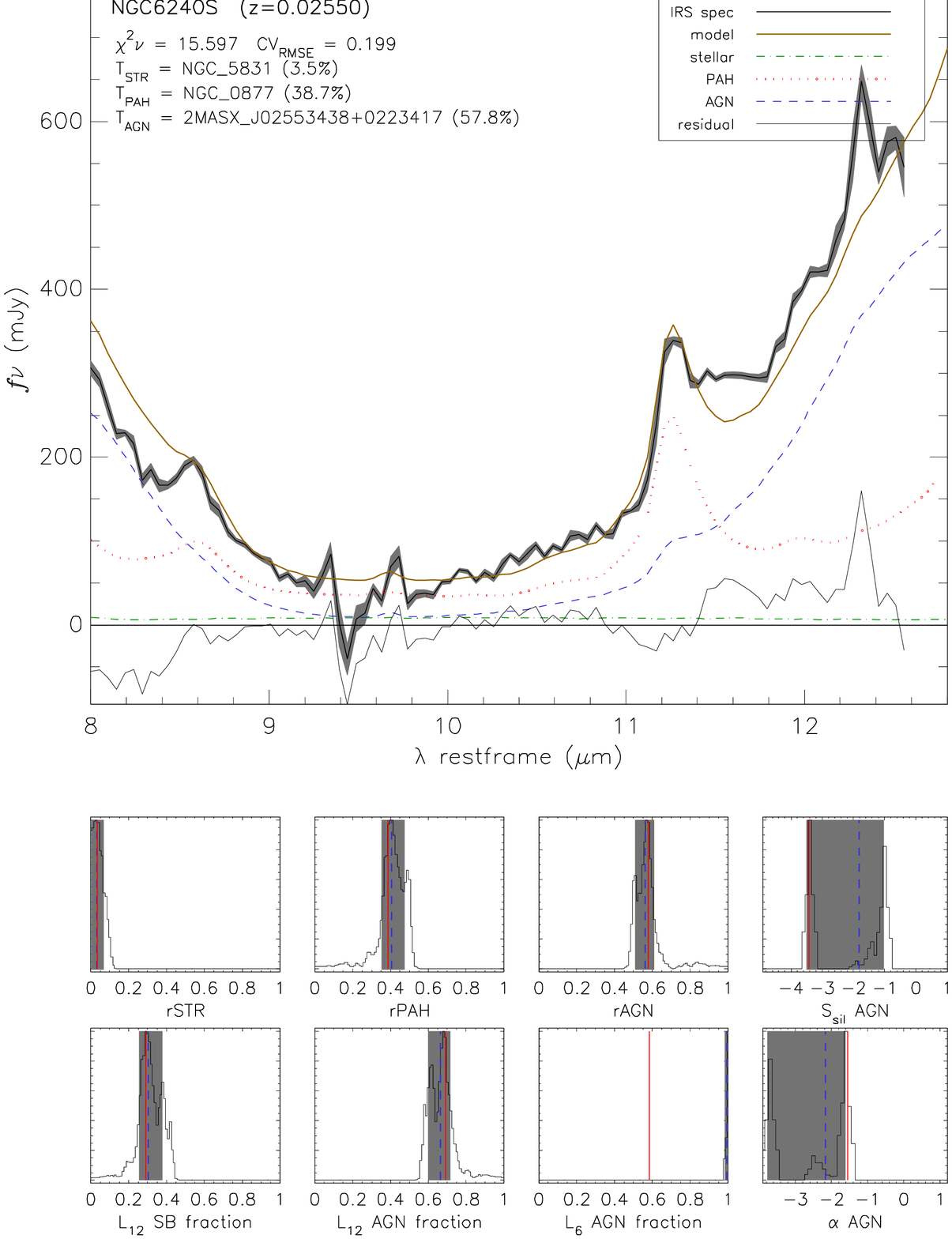}
        \caption{As Figure~\ref{fig:examplesdeblendIRS} but for the
          three IR-bright nuclei with deblendIRS fits with $\chi^2>10$.}

    \label{fig:deepsilicatesanddeblendIRS}
\end{figure*}

In objects with very deep silicate absorption features, the flux near
the silicate minimum 
at $9.7\,\mu$m is very weak and can be a factor of $\sim$100 lower
than in the wings at 8 and $13\,\mu$m.  The larger IRS aperture may
admit substantially more diffuse emission from the host galaxy than
the slits of the ground-based instruments, which can partially fill in
the silicate absorption minimum and change the shape of the apparent
silicate profile.  This effect can be seen in the comparison of the
IRS and ground-based TReCS spectra of NGC~4418 in \cite{Roche2015},
where the absorption feature in the IRS spectrum is broader than the profile measured with
the ground-based instrument TReCS because contributions from extra-nuclear emission flatten the
minimum and broaden the wings of the silicate profile whilst enhancing
the PAH emission.  The same effects can be seen in the fit to 
IRAS~08572+3915N in Fig.~\ref{fig:deepsilicatesanddeblendIRS}.  As more and higher quality ground-based spectra
become available, it may be possible to employ them as templates, but
the broader wavelength coverage available from space missions would
not be available.  In the same figure we also show the other two
IR-bright nuclei whose fits have $\chi^2>10$. 
IC~694 might suffer from the same problem with the silicate 
feature as IRAS~08572+3915N but it is not as extreme. Also the 
high value of $\chi^2$ is due to the relatively poor fits of the PAH
features. In
any case, both nuclei have small
$1\sigma$ uncertainties in the derived AGN $S_{\rm Sil}$ value due the above
mentioned small number of ``obscured AGN'' templates. For NGC~6240S the high value of
$\chi^2$ is likley caused by a bad atmospheric correction around $9.4-9.8\,\mu$m and at
the edges of the spectrum. We conclude that for most objects in our sample the IRS
templates are appropriate for the spectral decomposition of the nuclear mid-IR spectra.

\section{Comparison of  {\sc deblendIRS} results for {\it Spitzer}/IRS and high angular
resolution ground-based spectra}\label{appendix2}
\begin{figure*}
	\includegraphics[width=0.9\columnwidth]{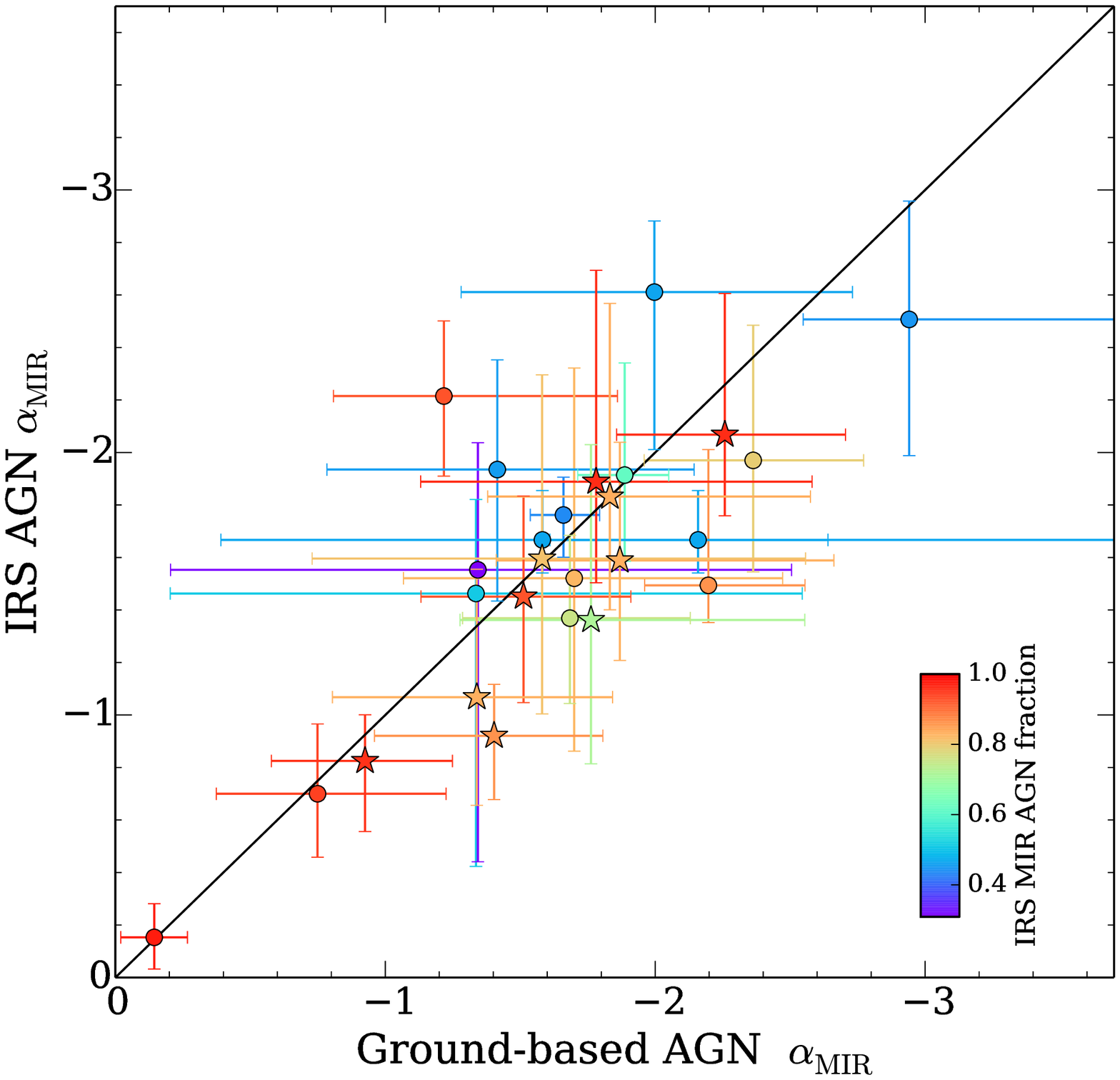}
	\includegraphics[width=0.9\columnwidth]{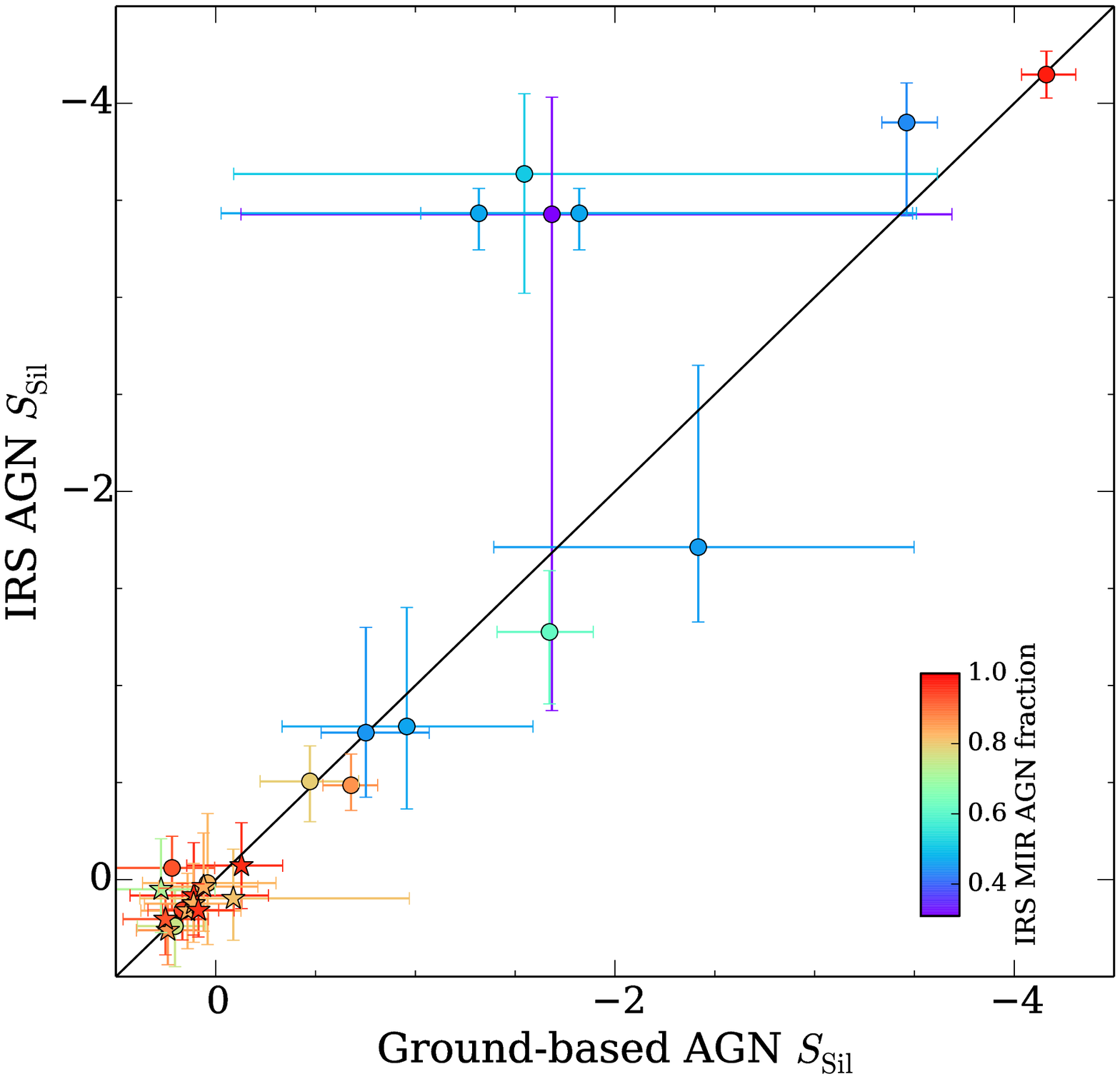}
        \caption{Comparison between the AGN mid-IR spectral index
          (left panel) and the AGN strength of the silicate feature (right
        panel) derived with {\sc deblendIRS} for the ground-based and
        {\it Spitzer}/IRS spectra. The  star symbols are the IR-weak
        quasars, whereas the circles are the IR-bright galaxies
        (including the IR-bright quasars). The straight line is the
        1:1 relation.}  
    \label{fig:comparisonalphasSSil}
\end{figure*}

In this appendix we compare the AGN $5-15\,\mu$m spectral index
$\alpha_{\rm MIR}$ and the strength of the silicate
feature $S_{\rm Sil}$ derived with {\sc deblendIRS} using {\it Spitzer}/IRS SL spectroscopy (3.7\,arcsec slit width) and
high-angular resolution ground-based spectroscopy (slit widths between
0.52 and 0.75\,arcsec) for the samples of IR-bright galaxies and
quasars. As done for the ground-based data, for the {\it Spitzer}/IRS
spectra we only fitted the approximate spectral range $8-12.5\,\mu$m
with {\sc deblendIRS}. We show the comparison between the derived AGN
$\alpha_{\rm MIR}$ and $S_{\rm Sil}$ in
Fig.~\ref{fig:comparisonalphasSSil}. As can be seen from this figure, the
agreement between the derived AGN properties is good within the derived
$1\sigma$ uncertainties. While the $1\sigma$ uncertainties of the derived AGN
$\alpha_{\rm MIR}$ and $S_{\rm Sil}$ for the quasars are similar for
the ground-based and IRS spectra, in some IR-bright nuclei, especially
those with low AGN fractional components, the IRS uncertainties are
slightly smaller probably due to the higher
S/N ratio of the IRS spectra. 
The most discrepant points correspond to some of the 
IR-bright nuclei with deep silicate features and/or for which the {\it
  Spitzer}/IRS spectroscopy could not isolate the individual nuclei (e.g., NGC~6240
and IRAS~14348$-$1447).

\bsp	
\label{lastpage}
\end{document}